\documentclass[fleqn,usenatbib]{mnras}

\usepackage{newtxtext,newtxmath}
\usepackage[T1]{fontenc}

\usepackage[british]{babel}

\usepackage[inline]{enumitem}

\DeclareRobustCommand{\VAN}[3]{#2}
\let\VANthebibliography\thebibliography
\def\thebibliography{\DeclareRobustCommand{\VAN}[3]{##3}\VANthebibliography}

\usepackage{graphicx}	% Including figure files
\usepackage{amsmath}	% Advanced maths commands
\usepackage{mathtools}
\usepackage[nameinlink]{cleveref}

\usepackage{pifont}% http://ctan.org/pkg/pifont
\newcommand{\cmark}{\ding{51}}%
\newcommand{\xmark}{\ding{55}}%

\usepackage{caption}
\usepackage{subcaption}
\usepackage{bm}

\newcommand{\sTheta}{\bm{\upTheta}} %set Theta
\newcommand{\e}{\mathrm{e}}       %upright e for Eulers number

\graphicspath{{Figures/}}

\title[Null-test-based Bayesian model validation]{A general Bayesian model-validation framework based on null-test evidence ratios, with an example application to global 21-cm cosmology}

\author[Sims et al.]{Peter H. Sims,$^{1}$\thanks{E-mail: psims3@asu.edu}
Steven G. Murray,$^{1,2}$
Judd D. Bowman,$^{1}$
John P. Barrett,$^{3}$
Rigel C. Cappallo,$^{3}$ \and
Colin J. Lonsdale,$^{3}$
Nivedita Mahesh,$^{4}$
Raul A. Monsalve,$^{1,5,6}$
Alan E. E. Rogers,$^{3}$
Titu Samson,$^{1}$ \and
and Akshatha K. Vydula$^{1}$
\\
$^{1}$School of Earth and Space Exploration, Arizona State University, Tempe, AZ 85287, USA\\
$^{2}$Scuola Normale Superiore (SNS), Piazza dei Cavalieri 7, I-56125 Pisa, PI, Italy\\
$^{3}$MIT Haystack Observatory, Westford, MA 01886-1299, USA \\
$^{4}$Cahill Center for Astronomy and Astrophysics, California Institute of Technology, Pasadena CA 91125, USA\\
$^{5}$Space Sciences Laboratory, University of California Berkeley,
Berkeley, CA 94720, USA \\
$^{6}$Facultad de Ingeniería, Universidad Católica de la Santísima Concepción, Alonso de Ribera 2850, Concepción, Chile
}

\date{Accepted XXX. Received YYY; in original form ZZZ}

\pubyear{2024}

\begin{document}
\label{firstpage}
\pagerange{\pageref{firstpage}--\pageref{lastpage}}
\maketitle

\begin{abstract}
    Comparing composite models for multi-component observational data is a prevalent scientific challenge. When fitting composite models, there exists the potential for systematics from a poor fit of one model component to be absorbed by another, resulting in the composite model providing an accurate fit to the data in aggregate but yielding biased a posteriori estimates for individual components.  We begin by defining a classification scheme for composite model comparison scenarios, identifying two categories: \textit{category I}, where models with accurate and predictive components are separable through Bayesian comparison of the unvalidated composite models, and \textit{category II}, where models with accurate and predictive components may not be separable due to interactions between components, leading to spurious detections or biased signal estimation. To address the limitations of \textit{category II} model comparisons, we introduce the Bayesian Null Test Evidence Ratio-based (BaNTER) validation framework. Applying this classification scheme and BaNTER to a composite model comparison problem in 21-cm cosmology, where minor systematics from imperfect foreground modelling can bias global 21-cm signal recovery, we validate six composite models using mock data. We show that incorporating BaNTER alongside Bayes-factor-based comparison reliably ensures unbiased inferences of the signal of interest across both categories, positioning BaNTER as a valuable addition to Bayesian inference workflows with potential applications across diverse fields.
\end{abstract}


%%%%%%%%%%%%%%%%%%%%%%%%%%%%%%%%%%%%%%%%%%%%%%%%%%
%%%%%%%%%%%%%%%%% BODY OF PAPER %%%%%%%%%%%%%%%%%%

%%%%%%%%%%%%%%%%%%%%%%%%%%%%%%%%%%%%%%%%%%%%%%%%%%
\section{Introduction}
\label{Sec:Intro}
%%%%%%%%%%%%%%%%%%%%%%%%%%%%%%%%%%%%%%%%%%%%%%%%%%

Comparing competing models for observational data is a cornerstone of scientific research. In the absence of a definitive model, one must account for uncertainties in both the model parameters and the model's description of the data to draw robust inferences. Ignoring the latter can lead to biased\footnote{We use `bias' to refer to systematic errors in parameter estimates that result from model incompleteness or misspecification, where the model does not adequately capture the true data-generating process.} parameter estimates and underestimated uncertainties (e.g. \citealt{2020MNRAS.492...22S}; hereafter, S20).

Bayesian model comparison provides a statistically consistent means by which both model and parameter uncertainty can be accounted for (e.g. \citealt{1935PCPS...31..203J, KandR}). In its most general form, Bayesian model comparison calculates the posterior odds between models for the data. This is calculated from two factors: the prior odds of the models (our initial belief in the models) and the Bayes factor (the relative consistency of the models' fits to the data, weighted by parameter priors).

If there is no a priori information favouring any model, the prior odds are unity, and the Bayes factor and posterior odds are equivalent. In this limit, and assuming comparison of non-composite models of a data set that contains a single signal component, Bayes-factor-based model comparison (hereafter, BFBMC) provides a means of calculating the more predictive models for that signal\footnote{We use `predictivity' to describe a model's statistical consistency with the data over its prior volume (equivalent to its Bayesian evidence).}.

A more challenging problem arises when comparing models for data containing multiple signals, one of which is of primary interest, while the rest represent nuisance structures. Examples in astrophysics include:
\begin{itemize}
    \item modelling of data containing the Cosmic Microwave Background and nuisance foreground emission at microwave frequencies (e.g. \citealt{2020A&A...641A...6P}),
    \item modelling of biomarker signatures in exoplanet transmission spectra in the presence of potential instrumental systematics (e.g. \citealt{2023ApJ...956L..13M, 2023ApJ...948L..11M}), and
    \item the modelling of data containing 21-cm radiation emitted by neutral hydrogen at Cosmic Dawn (e.g. \citealt{2018Natur.555...67B}) in the presence of nuisance radio frequency foreground emission (e.g. \citealt{2019MNRAS.489.4007S, 2020PASP..132f2001L}) and potential instrumental systematics (e.g.  \citealt{2018Natur.564E..32H, 2019ApJ...874..153B, 2019ApJ...880...26S}; S20; \citealt{2021MNRAS.502.4405B}).
\end{itemize}

In such contexts, unlike with models of data sets measuring a single signal, there is a potential for the set of composite models under consideration to include models that are predictive of the data in aggregate but that can only obtain accurate fits to the data with biased component model fits. While BFBMC can determine the most predictive models, it cannot distinguish between models that provide accurate aggregate fits but biased component model fits. Thus, in this case, a model's inclusion in the preferred set, as determined by BFBMC, is necessary but not sufficient for ensuring accurate recovery of the signal components.

The above scenario can be made more concrete with the following example. Consider a data set $\bm{D} = \bm{S}_\mathrm{a} + \bm{S}_\mathrm{b} + \bm{n}$, where $\bm{S}_\mathrm{a}$ is the primary signal of interest (SOI), $\bm{S}_\mathrm{b}$ is a nuisance signal\footnote{More generally, the data may contain $N$ nuisance signals that contribute structure to the data. For the purpose of forward modelling the data, these may be modelled individually; however, for this work we treat their sum as a single nuisance signal and the sum of the models for these signals as a single nuisance model.}, and $\bm{n}$ is noise. Furthermore, we have three competing composite models for the data $\bm{M}_{1\mathrm{c}} = \bm{M}_\mathrm{1a} + \bm{M}_\mathrm{1b}$, $\bm{M}_{2\mathrm{c}} =  \bm{M}_\mathrm{2a} + \bm{M}_\mathrm{2b}$ and $\bm{M}_{3\mathrm{c}} =  \bm{M}_\mathrm{3a} + \bm{M}_\mathrm{3b}$, such that each is the linear sum of two components that are intended to describe signals $\bm{S}_\mathrm{a}$ and $\bm{S}_\mathrm{b}$, respectively. Let us assume that $\bm{M}_{1\mathrm{c}}$ and $\bm{M}_{2\mathrm{c}}$ are comparably predictive models for the data (such that they fit the data equally well over their respective prior volumes; e.g. \citealt{2008ConPh..49...71T}) and both are significantly more predictive than $\bm{M}_{3\mathrm{c}}$. However, $\bm{M}_{1\mathrm{c}}$ accurately fits the data when its components accurately describe their signal counterparts. In contrast, accurate fits to the data with $\bm{M}_{2\mathrm{c}}$ are only\footnote{Here, we are referring to a model with components with which it is \textit{impossible} to accurately describe their signal counterparts at the signal-to-noise level of the data but which in combination can describe the data in aggregate. It does not, for example, refer to a model for which the posteriors of the model parameters in a fit to the data are consistent with those parameters in nature, regardless of whether there are degeneracies between parameters.} achievable through an inaccurate fit of $\bm{M}_\mathrm{2b}$ to $\bm{S}_\mathrm{b}$ with the residual structure in $\bm{S}_\mathrm{b}$ that is unmodelled by $\bm{M}_\mathrm{2b}$ being absorbed by $\bm{M}_\mathrm{2a}$.

Applying BFBMC to such a set of models and data, one will identify $\bm{M}_{1\mathrm{c}}$ and $\bm{M}_{2\mathrm{c}}$ as more predictive models of the data than $\bm{M}_{3\mathrm{c}}$; however, one will not be able to infer which of the models ($\bm{M}_{1\mathrm{c}}$ and $\bm{M}_{2\mathrm{c}}$) is more likely to yield unbiased inferences about the structure of $\bm{S}_\mathrm{a}$. Thus, using BFBMC to compare $\bm{M}_{1\mathrm{c}}$, $\bm{M}_{2\mathrm{c}}$ and $\bm{M}_{3\mathrm{c}}$ and weighting our models according to their relative Bayes factor will result in a spuriously high probability being assigned to biased parameter inferences from $\bm{M}_{2\mathrm{c}}$.

Breaking this degeneracy in the posterior model probabilities of $\bm{M}_{1\mathrm{c}}$ and $\bm{M}_{2\mathrm{c}}$ is possible if one has a priori knowledge of the model components that suggests that they are not equally credible\footnote{We use `credibility' to refer to the degree of belief placed in a model.}
descriptions of their signal counterparts in the data. One way to obtain this prior information is through a null test comparing fits of the nuisance and composite models to SOI-free validation data, $\bm{D}_\mathrm{v}$. If the composite model is preferred over the nuisance model, it indicates the nuisance model is insufficiently predictive for unbiased SOI inference, and the null test is failed.

In this paper, we present BaNTER, a Bayesian Null Test Evidence Ratio-based validation framework. BaNTER calculates the evidence ratio (Bayes factor) between nuisance and composite models for SOI-free validation data. These validation results are used to make an a priori assessment of the probability that unbiased estimates of the SOI are obtainable with the composite model. We combine this with an analysis of the a posteriori probability of the data from BFBMC in a validated posterior-odds-based model comparison methodology.

We demonstrate an application of this approach to the problem of inferring accurate and precise estimates of the cosmological signal in global 21-cm cosmology mock data. In this test case, despite one of the models under consideration being the generative model for the mock data, using BFBMC alone does not reliably downweight biased alternatives. Consequently, it leads to biased 21-cm signal estimation. In contrast, we demonstrate that a BaNTER validated posterior-odds-based methodology enables unbiased estimates of the 21-cm SOI to be obtained.

In this work, we emphasize a detailed explanation of the BaNTER validation methodology. We use a test case with a compact set of models to demonstrate its efficacy and to highlight salient features. In Sims et al. (in preparation), we apply the BaNTER framework to a broader comparison of data models that have been applied to 21-cm signal estimation in the literature (\citealt{2018Natur.555...67B, 2018Natur.564E..32H, 2023MNRAS.521.3273S}).

The remainder of the paper is organised as follows. In \Cref{Sec:SIMVAMPTheory}, we introduce our composite model validation methodology. In \Cref{Sec:Results}, we apply it to a comparison of competing models for beam-factor-chromaticity-corrected global 21-cm signal data. We discuss the efficacy and limitations of the validation methodology, and potential improvements in \Cref{Sec:Discussion}. In \Cref{Sec:Conclusions} we provide a summary and discuss avenues for future work.

%%%%%%%%%%%%%%%%%%%%%%%%%%%%%%%%%%%%%%%%%%%%%%%%%%
\section{Null-test-based Bayesian model validation}
\label{Sec:SIMVAMPTheory}
%%%%%%%%%%%%%%%%%%%%%%%%%%%%%%%%%%%%%%%%%%%%%%%%%%

The logical possibility exists for composite models that are predictive of the data in aggregate to obtain accurate fits with biased component models. However, in the absence of informative model priors one does not know whether models of this type are included in the set of models under consideration. When they are, treating these models as a priori equally probable will result in a degeneracy in the space of a posteriori model probabilities between these models and those that we wish to identify (those that are both predictive of the data in aggregate and have predictive component models). The purpose of the model validation framework introduced in this section is to identify when a set of competing composite models exhibits this degeneracy and to remove it by identifying and downweighting models that can only achieve accurate fits to the data using biased fits of its component models.

In \Cref{Sec:BayesianInference}, we begin by introducing the principles of Bayesian inference, upon which our model validation and posterior-odds-based model comparison method is based. In \Cref{Sec:InformativeModelPriors}, we introduce a classification scheme of model comparison scenarios based on whether they exhibit the aforementioned degeneracy between preferred models for the data as judged by BFBMC. This also serves to introduce the model set notation used in the remainder of the paper. In \Cref{Sec:BaNTER} we present BaNTER validation that is designed to eliminate this degeneracy, enabling unbiased inferences of the SOI.

%%%%%%%%%%%%%%%%%%%%%%%%%%%%%%%%%%%%%%%%%%%%%%%%%%
\subsection{Bayesian inference}
\label{Sec:BayesianInference}
%%%%%%%%%%%%%%%%%%%%%%%%%%%%%%%%%%%%%%%%%%%%%%%%%%

Bayesian inference provides a consistent approach to estimate a set of parameters, $\sTheta$, from a model, $\bm{M}$, given a set of data, $\bm{D}$. Using Bayes' theorem we can write the posterior probability density of the parameters of the model as:
\begin{equation}
\label{Eq:BayesEqn}
\mathcal{P}(\sTheta\vert\bm{D},\bm{M}) = \dfrac{\mathcal{P}(\bm{D}\vert\sTheta,\bm{M})\ \mathcal{P}(\sTheta\vert \bm{M})}{\mathcal{P}(\bm{D}\vert \bm{M})} \ .
\end{equation}
Here, $\mathcal{P}(\bm{D}\vert\sTheta,\bm{M}) \equiv \mathcal{L}(\sTheta)$ is the likelihood of the data, $\mathcal{P}(\sTheta\vert \bm{M}) \equiv \pi(\sTheta)$ is the prior probability density of the parameters and $\mathcal{P}(\bm{D}\vert \bm{M}) \equiv \mathcal{Z} = \int\mathcal{L}(\sTheta)\pi(\sTheta)\mathrm{d}^{n}\sTheta$ is the Bayesian evidence, where $n$ is the dimensionality of the parameter space.

When, rather than a single definitive model, one has a set of models for the data, $\bm{\mathcal{M}} = \{\bm{M}_{1}, \bm{M}_{2}, \cdots, \bm{M}_{N}\}$, preferred models for the data can be determined from their marginal probabilities. Using Bayes' theorem at the model level, the marginal probability of a model $\bm{M}_{i}$, drawn from $\bm{\mathcal{M}}$, is given by:
\begin{equation}
\label{Eq:BayesEqnForModels}
\mathcal{P}(\bm{M}_{i} \vert \bm{D},\bm{\mathcal{M}}) =  \frac{\mathcal{P}(\bm{D} \vert \bm{M}_{i},\bm{\mathcal{M}}) \mathcal{P}(\bm{M}_{i} \vert \bm{\mathcal{M}})}{\mathcal{P}(\bm{D}\vert \bm{\mathcal{M}})} \ .
\end{equation}
Here, $\mathcal{P}(\bm{D} \vert \bm{\mathcal{M}}) = \sum_{k=1}^{N} \mathcal{P}(\bm{D}\vert \bm{M}_{k},\bm{\mathcal{M}}) \mathcal{P}(\bm{M}_{k} \vert \bm{\mathcal{M}})$ is the marginal probability of the data over the models and their parameters, $\mathcal{P}(\bm{D} \vert \bm{M}_{i},\bm{\mathcal{M}})$ is the Bayesian evidence of $\bm{M}_{i}$ and $\mathcal{P}(\bm{M}_{i} \vert \bm{\mathcal{M}})$ is the probability of $\bm{M}_{i}$ prior to analysing the data. For brevity, we leave the conditioning of the probability densities on $\bm{\mathcal{M}}$ implicit going forward.

For the purpose of comparing two specific models for the data, $\bm{M}_{i}$ and $\bm{M}_{j}$, drawn from $\bm{\mathcal{M}}$, one can calculate their posterior odds:
\begin{align}
\label{Eq:RgivenData}
\underbrace{\mathcal{R}_{ij}}_{\text{Posterior odds}} &= \dfrac{\mathcal{P}(\bm{M}_{i}\vert\bm{D})}{\mathcal{P}(\bm{M}_{j}\vert\bm{D})} \\ \nonumber
&= \dfrac{\mathcal{P}(\bm{D}\vert \bm{M}_{i})\mathcal{P}(\bm{M}_{i})}{\mathcal{P}(\bm{D}\vert \bm{M}_{j})\mathcal{P}(\bm{M}_{j})} \\ \nonumber
&= \underbrace{\mathcal{B}_{ij}}_{\text{Bayes factor}} \underbrace{\dfrac{\mathcal{P}(\bm{M}_{i})}{\mathcal{P}(\bm{M}_{j})}}_{\text{Prior odds}} \ .
\end{align}
Here, $\mathcal{P}(\bm{M}_{i}\vert\bm{D})$ is given by \Cref{Eq:BayesEqnForModels}, the Bayes Factor, $\mathcal{B}_{ij}$, is the posterior odds of the data given $\bm{M}_{i}$ and $\bm{M}_{j}$, $\mathcal{P}(\bm{D}\vert \bm{M}_{i}) \equiv \mathcal{Z}_{i}$ and $\mathcal{P}(\bm{D}\vert \bm{M}_{j}) \equiv \mathcal{Z}_{j}$ are the marginal likelihoods of the data in models $\bm{M}_{i}$ and $\bm{M}_{j}$, respectively, and $\mathcal{P}(\bm{M}_{i})/\mathcal{P}(\bm{M}_{j})$ is the ratio of the prior probabilities of the two models before any conclusions have been drawn from the data.

As the model-prior-weighted average of the likelihood over the parameters priors, the marginal probability of a model is larger if the model is probable a priori and more of its parameter space is likely given the data; it is smaller for a model that is improbable a priori or if large areas of its parameter space have low likelihood values, even if the likelihood function is very highly peaked. It thus represents an updating of one's prior credence in the model, given the data, and automatically incorporates an `Occam penalty' against a more complex theory with a broad parameter space. As such, in absence of an a priori reason to prefer it over a simpler alternative, it will be favoured only if it is significantly better at explaining the data.

In this work, we use the definitions in \Cref{Tab:aPosterioriPreference} to map between the posterior odds of two models and qualitative terms describing the relative preference for one over the other. When we have no information that lends additional credibility to one model over the other in advance of analysing the data, we set the prior odds to unity. In this case, the posterior odds are equal to the Bayes factor between the models ($\mathcal{R}_{ij}=\mathcal{B}_{ij}$), and for the purpose of defining our qualitative descriptions the Bayes factor between models takes the place of the posterior odds. We have chosen boundaries between our qualitative descriptions such that in this limit they map to those given by \cite{KandR} for the preference of one model over another.

\begin{table}
\caption{
    Standards of a posteriori model preference for model $\bm{M}_{i}$ over $\bm{M}_{j}$ as a function of $\ln({\mathcal{R}_{ij}})$, the log of the posterior odds between the two models. If one has no a priori information that makes any one of the models more probable than the others, the prior odds are unity and the Bayes factor and the posterior odds between models are equivalent. In this situation, we use the same qualitative descriptions for matching ranges of $\ln({\mathcal{B}_{ij}})$.
}
\centerline{
    \begin{tabular}{l l l }
        \hline
        $\ln({\mathcal{R}_{ij}})$ & Odds in favour of $\bm{M}_{i}$ & Preference for $\bm{M}_{i}$ \\
        \hline
        $0 \le \ln({\mathcal{R}_{ij}}) < 1$ & 1--3 & Weak \\
        $1 \le \ln({\mathcal{R}_{ij}}) < 3$ & 3--20 & Moderate \\
        $3 \le \ln({\mathcal{R}_{ij}}) < 5$ & 20--150 & Strong \\
        $\ln({\mathcal{R}_{ij}}) \ge 5$ & $>150$  & Decisive \\
        \hline
    \end{tabular}
}
\label{Tab:aPosterioriPreference}
\end{table}

\subsection{Classifying model comparison scenarios}
\label{Sec:InformativeModelPriors}

\subsubsection{Motivation}
\label{Sec:Motivation}

Suppose we wish to describe a data set of the form given in the introduction:
\begin{equation}
    \label{Eq:Data}
    \bm{D} = \bm{S}_\mathrm{a} + \bm{S}_\mathrm{b} + \bm{n} \ ,
\end{equation}
with $\bm{S}_\mathrm{a}$ the SOI. Consider two sets of models for describing $\bm{S}_\mathrm{a}$ and $\bm{S}_\mathrm{b}$:
\begin{align}
    \label{Eq:CurlyMa}
    \bm{\mathcal{M}}_{\mathrm{a}}
    &= \{\bm{M}_{1\mathrm{a}}, \cdots, \bm{M}_{N_\mathrm{a}\mathrm{a}} \} = \{\bm{M}_{j\mathrm{a}} \mid j \in [1, N_\mathrm{a}]\} \
\end{align}
and
\begin{equation}
    \label{Eq:CurlyMb}
    \bm{\mathcal{M}}_{\mathrm{b}} = \{\bm{M}_{k\mathrm{b}} \mid k \in [1, N_\mathrm{b}]\} \ ,
\end{equation}
respectively. Additionally, let,
\begin{equation}
    \label{Eq:CurlyMc}
    \bm{\mathcal{M}}_{\mathrm{c}} = \{\bm{M}_{i\mathrm{c}} \mid i \in [1, N_\mathrm{c}]\} \ ,
\end{equation}
denote the set of composite models obtained by summing pairs of models, one from $\bm{\mathcal{M}}_{\mathrm{a}}$ and one from $\bm{\mathcal{M}}_{\mathrm{b}}$\footnote{More generally, one could consider data and models of the form $\bm{D} = f(\bm{S}_\mathrm{a} + \bm{S}_\mathrm{b}) + \bm{n}$ and $\bm{M}_{i\mathrm{c}} = f_{M}(\bm{M}_{i\mathrm{a}} + \bm{M}_{i\mathrm{b}})$, respectively. Here, $f(.)$ is a function (such as an instrumental transfer function) that operates on the signal components and $f_{M}(.)$ is our model for that function. The approach discussed in this work generalises straightforwardly to this scenario. For notational simplicity, we treat $f$ as the identity function in what follows.}. Here, $N_\mathrm{a}$, $N_\mathrm{b}$ and $N_\mathrm{c}$ are the number of models in the three sets.

In the introduction we considered three competing composite models ($\bm{M}_{1\mathrm{c}}$, $\bm{M}_{2\mathrm{c}}$ and $\bm{M}_{3\mathrm{c}}$), with $\bm{M}_{1\mathrm{c}}$ and $\bm{M}_{2\mathrm{c}}$ being comparably predictive of the data and both being significantly more predictive than $\bm{M}_{3\mathrm{c}}$. However, $\bm{M}_{1\mathrm{c}}$ accurately fits the data when its components accurately describe their signal counterparts, while accurate fits to the data with $\bm{M}_{2\mathrm{c}}$ are only achievable through an inaccurate fit of $\bm{M}_\mathrm{2a}$ to $\bm{S}_\mathrm{a}$, with the residual structure in $\bm{S}_\mathrm{a}$ that is unmodelled by $\bm{M}_\mathrm{2a}$ being absorbed by $\bm{M}_\mathrm{2b}$. BFBMC will separate $\bm{M}_{1\mathrm{c}}$, which we aim to identify, from the significantly lower predictivity composite model, $\bm{M}_{3\mathrm{c}}$. However, it will not yield a preference for $\bm{M}_{1\mathrm{c}}$ over $\bm{M}_{2\mathrm{c}}$, due to their comparable predictivities for the data in aggregate. Thus, in this case, BFBMC alone is sufficient for identifying the most predictive composite models for the data but not for uniquely identifying $\bm{M}_{1\mathrm{c}}$ as the only composite models that additionally has component models that accurately describe their signal counterparts.

Counterposing this example, the logical possibility also exists for the set of models under consideration to be composed exclusively of composite models that are predictive of the data and have components that accurately describe their signal counterparts (such as $\bm{M}_{1\mathrm{c}}$), as well as composite models that are significantly less predictive of the data (such as $\bm{M}_{3\mathrm{c}}$). In this simpler case, BFBMC alone is sufficient for identifying both the most predictive models for the data \textit{and} the most predictive component models for the signals, rather than the former alone.

Let us label these two cases as \textit{category II} (challenging) and \textit{category I} (simple) composite model comparison scenarios, respectively. In the \textit{category I} scenario, model validation is incidental, while in the \textit{category II} scenario, it is essential. To determine under what conditions model validation is required, it is necessary to understand the properties of the composite and component models that lead to \textit{category I} and \textit{II} model comparison scenarios. In the remainder of this section, we analyse the logical relations between the sets of models in $\bm{\mathcal{M}}_{\mathrm{a}}$, $\bm{\mathcal{M}}_{\mathrm{b}}$ and $\bm{\mathcal{M}}_{\mathrm{c}}$ and classify under what conditions these two scenarios arise.

\subsubsection{Categorising models by their predictivity and accuracy}
\label{Sec:ModelPredictivityAndAccuracy}

To identify when one is dealing with a \textit{category I} or \textit{II} model comparison problem, and thus whether model validation is required, let us classify the models in $\bm{\mathcal{M}}_{\mathrm{a}}$, $\bm{\mathcal{M}}_{\mathrm{b}}$, and $\bm{\mathcal{M}}_{\mathrm{c}}$ according to whether they meet a given predictivity and accuracy\footnote{We use `accuracy' to refer to the statistical consistency of $\bm{M}(\overline{\sTheta})$ with the data, with $\overline{\sTheta}$ the median parameter sample for the posterior.} threshold. This will further serve to construct a notation for distinguishing the desired set of composite models from the remainder that are either not predictive of the data in aggregate, or predictive of the data in aggregate but composed of components that are not predictive of their signal counterparts.

We will use predictivity and accuracy criteria to classify models in each of $\bm{\mathcal{M}}_{\mathrm{a}}$, $\bm{\mathcal{M}}_{\mathrm{b}}$, and $\bm{\mathcal{M}}_{\mathrm{c}}$ into pairs of complementary subsets. Each pair will consist of one subset containing the preferred models that meet the predictivity and accuracy criteria for the signal component(s) they are intended to describe, and another containing the disfavoured models that do not meet these criteria.

We begin by defining the predictivity and accuracy criteria in the context of the composite models in $\bm{\mathcal{M}}_{\mathrm{c}}$, and will return to discuss their application to $\bm{\mathcal{M}}_{\mathrm{a}}$ and $\bm{\mathcal{M}}_{\mathrm{b}}$ afterwards.

For the composite models in $\bm{\mathcal{M}}_{\mathrm{c}}$, we define the subsets containing the preferred and disfavoured models as $C$ and $\overline{C}$, respectively.

Given the dataset $\bm{D}$ and the noise covariance matrix $\mathbfss{N}$, we denote by $\mathcal{Z}_{\mathrm{max}}$ the Bayesian evidence of the most predictive model for the data, $\bm{M}_{\mathrm{c,max}}$, and by $\mathcal{B}_{i\mathrm{max}}$ the Bayes factor between the $i$th model, $\bm{M}_{i\mathrm{c}}$, and $\bm{M}_{\mathrm{c,max}}$, such that
\begin{equation}
    \label{Eq:Bimax}
    \mathcal{B}_{i\mathrm{max}} = \frac{\mathcal{Z}_{i}}{\mathcal{Z}_{\mathrm{max}}} \ .
\end{equation}
We also define the median a posteriori likelihood of $\bm{M}_{i\mathrm{c}}$ as $\log(\overline{\mathcal{L}}_{i})$. Writing the data likelihood $\mathcal{L}(\bm{r}_{i\mathrm{c}}(\sTheta_{i\mathrm{c}}))$ as a function of the residual vector, $\bm{r}_{i\mathrm{c}}(\sTheta_{i\mathrm{c}}) = [\bm{D} - \bm{M}_{i\mathrm{c}}(\sTheta_{i\mathrm{c}})]$ (see \Cref{Sec:DataLikelihood}), the likelihood distribution for an ideal model (one that describes the data perfectly, excluding noise) can be sampled by substituting $\bm{r}_{i\mathrm{c}}(\sTheta_{i\mathrm{c}})$ in the likelihood expression with noise realizations drawn from the covariance matrix $\mathbfss{N}$. We denote this ideal model likelihood distribution as $\mathcal{L}_\mathrm{noise}$. Finally, we define the model's accuracy parameter as the logarithm of the ratio of the median a posteriori likelihood of $\bm{M}_{i\mathrm{c}}$ to the ideal model likelihood distribution:
\begin{equation}
    \label{Eq:lambdaLR}
    \lambda_{i} = \log\left( \frac{\overline{\mathcal{L}}_{i}}{\mathcal{L}_\mathrm{noise}} \right) \ .
\end{equation}
When the distribution of $\lambda_{i}$ is consistent with zero, this implies $\overline{\mathcal{L}}_{i}$ is comparable to typical values of $\mathcal{L}_\mathrm{noise}$ and $\bm{M}_{i\mathrm{c}}$ is accurate. In contrast, when most of the probability mass of $\lambda_{i}$ is negative, $\bm{M}_{i\mathrm{c}}$ is comparably inaccurate.

Qualitatively, we define a predictive and accurate composite model as one with a Bayesian evidence equal to or comparable to the most predictive model, and a fit likelihood that is credibly drawn from the ideal model likelihood distribution. Quantitatively, we classify $\bm{M}_{i\mathrm{c}}$ as an element of $C$ if it satisfies the following predictivity and accuracy conditions:
\begin{enumerate}
    \item \textbf{Predictivity condition:}
    \begin{equation}
        \label{Eq:PredictivityCondition}
        \log(\mathcal{B}_{i\mathrm{max}}) \ge \log(\mathcal{B}_\mathrm{threshold}) ;
    \end{equation}
    \item \textbf{Accuracy condition:}
    \begin{equation}
        \label{Eq:AccuracyCondition}
        Q_{q_\mathrm{threshold}}(\lambda_{i}) \ge 0 \ .
    \end{equation}
\end{enumerate}
Here, $Q(.)$ is the quantile (or inverse cumulative distribution) function, defined such that for a random variable $X$, $Q_q(X)$ is the value of $x$ such that $P(X \leq x) = q$. The closer $q_\mathrm{threshold}$ is to unity, the further $\overline{\mathcal{L}}_i$ can fall towards the lower end of the $\mathcal{L}_\mathrm{noise}$ distribution while still being classified as an element of $C$.

In this work, we define $\log(\mathcal{B}_\mathrm{threshold}) = -3$ and $q_\mathrm{threshold} = 0.999$. Condition (i) guarantees that the odds in favour of $\bm{M}_{\mathrm{c,max}}$ over $\bm{M}_{i\mathrm{c}}$ are at most 20-to-1 (see \Cref{Tab:aPosterioriPreference}), meaning that $\bm{M}_{i\mathrm{c}}$ is among the most predictive of the composite models under consideration. However, this condition alone does not address situations in which none of the models can accurately describe the data. Condition (ii) addresses this by ensuring that the residuals of the fitted model are consistent with the noise in the data. Specifically, $\bm{M}_{i\mathrm{c}}$ will fail condition (ii) only if its median posterior likelihood is exceeded by 99.9\% of the probability mass of the $\mathcal{L}_\mathrm{noise}$ distribution. If a composite model does not fulfil both of these conditions, it is classified as an element of $\overline{C}$.

For the purposes of model categorisation in \Cref{Sec:CategoryI,Sec:CategoryII}, we assume that $\bm{\mathcal{M}}_{\mathrm{c}}$ contains at least one model that is an element of $C$. When this is known a priori, one can directly assess whether the models in $\bm{\mathcal{M}}_{\mathrm{c}}$ provide predictive and accurate descriptions of the data using condition (i). However, if this is uncertain, one can test whether condition (ii) holds for the models in $\bm{\mathcal{M}}_{\mathrm{c}}$, to validate that this is the case.

Given $\bm{D}$, $\mathbfss{N}$, and the set of composite models for the data $\bm{\mathcal{M}}_{\mathrm{c}}$, there is no barrier to calculating the quantities required to assess whether a model fulfils conditions (i) and (ii), and thus to sort the models a posteriori into the sets $C$ and $\overline{C}$. In contrast, direct access to equivalent datasets containing only $\bm{S}_\mathrm{a}$ and $\bm{S}_\mathrm{b}$ may not be available, potentially necessitating more approximate, simulation-based methods to assess the predictivity of the models in $\bm{\mathcal{M}}_{\mathrm{a}}$ and $\bm{\mathcal{M}}_{\mathrm{b}}$. In either case, one can nevertheless define equivalent subsets of models in $\bm{\mathcal{M}}_{\mathrm{a}}$ and $\bm{\mathcal{M}}_{\mathrm{b}}$ to complete the conceptual framework\footnote{We will discuss how the membership of these sets can be inferred from the combined results of model validation and BFBMC later in the text.}. We classify the models in these sets according to whether they are predictive and accurate descriptions of their corresponding signal components, using the same criteria described for models in $\bm{\mathcal{M}}_{\mathrm{c}}$, but with the update that the signal component being fitted is exclusively $\bm{S}_\mathrm{a}$ or $\bm{S}_\mathrm{b}$, respectively. With this update, we define the models in $\bm{\mathcal{M}}_{\mathrm{a}}$ and $\bm{\mathcal{M}}_{\mathrm{b}}$ that satisfy their corresponding predictivity and accuracy conditions as elements of the sets $A$ and $B$, respectively, and their logical complements as elements of $\overline{A}$ and $\overline{B}$, respectively.

The predictivity and accuracy conditions given in \Cref{Eq:PredictivityCondition,Eq:AccuracyCondition} are chosen for their generality and because the quantities required to assess them are readily available, or can be easily computed, from the inputs and outputs of a Bayesian model comparison analysis using nested sampling (see \Cref{Sec:ComputationalTechniques}).

\subsubsection{Component Model Permutations and the Resulting Composites}
\label{Sec:ComponentModelPermutations}

As a stepping stone to categorising the models in $\bm{\mathcal{M}}_{\mathrm{c}}$ based on the categorisation of their component models, we consider the possible permutations of the sets $A$, $\overline{A}$, $B$, and $\overline{B}$ that can be used to construct composite models in $C$, and $\overline{C}$. Let us use arithmetic multiplication to represent the logical AND operation, such that $AB$ describes a pair of models, $\bm{M}_{j\mathrm{a}}$ and $\bm{M}_{k\mathrm{b}}$, drawn from $A$ and $B$, respectively. Composite models in $C$ and $\overline{C}$ are formed by combining pairs of models, one from $\bm{\mathcal{M}}_{\mathrm{a}}$ and one from $\bm{\mathcal{M}}_{\mathrm{b}}$. These pairs can be drawn from four permutations of $A$, $\overline{A}$, $B$, and $\overline{B}$, which we can write as: $AB$, $A\overline{B}$, $\overline{A}B$, and $\overline{A}~\overline{B}$.

An inaccurate or non-predictive composite model (i.e., an element of $\overline{C}$) can be constructed by combining models for $\bm{S}_\mathrm{a}$ and $\bm{S}_\mathrm{b}$ where at least one, and potentially both, models are inaccurate. This results in three permutations that can form composite models in $\overline{C}$, namely: $A\overline{B}~\overline{C}$, $\overline{A}B\overline{C}$, and $\overline{A}~\overline{B}~\overline{C}$. We note that a composite model composed of accurate and predictive components (i.e., when both $\bm{S}_\mathrm{a}$ and $\bm{S}_\mathrm{b}$ are represented by accurate and predictive models) will necessarily be accurate and predictive of the data in aggregate, and thus an element of $C$. Therefore, the combination $AB\overline{C}$ is not a valid permutation, and the corresponding set is empty.

In contrast, a composite model that can accurately describe and is predictive of the data in aggregate (and thus belongs to the set $C$) can be constructed by combining models for $\bm{S}_\mathrm{a}$ and $\bm{S}_\mathrm{b}$ where both are accurate, or when either or both are inaccurate, subject to the condition that any inaccuracies in one model can be absorbed by the other, such that the composite model remains accurate overall. Therefore, all four permutations can be used to construct composite models in $C$, namely: $ABC$, $A\overline{B}C$, $\overline{A}BC$, and $\overline{A}~\overline{B}C$.

To aid intuition for these set definitions their interpretation in the context of 21-cm cosmology is discussed in \Cref{Sec:EulerSets21-cmCosmologyExample}.

\subsubsection{Category I Model Comparison}
\label{Sec:CategoryI}

\begin{figure*}
	\centerline{
	\begin{subfigure}[t]{0.5\textwidth}
        \caption{\textit{Category I}}
        \label{Fig:EulerDiagram1}
        \includegraphics[width=\textwidth]{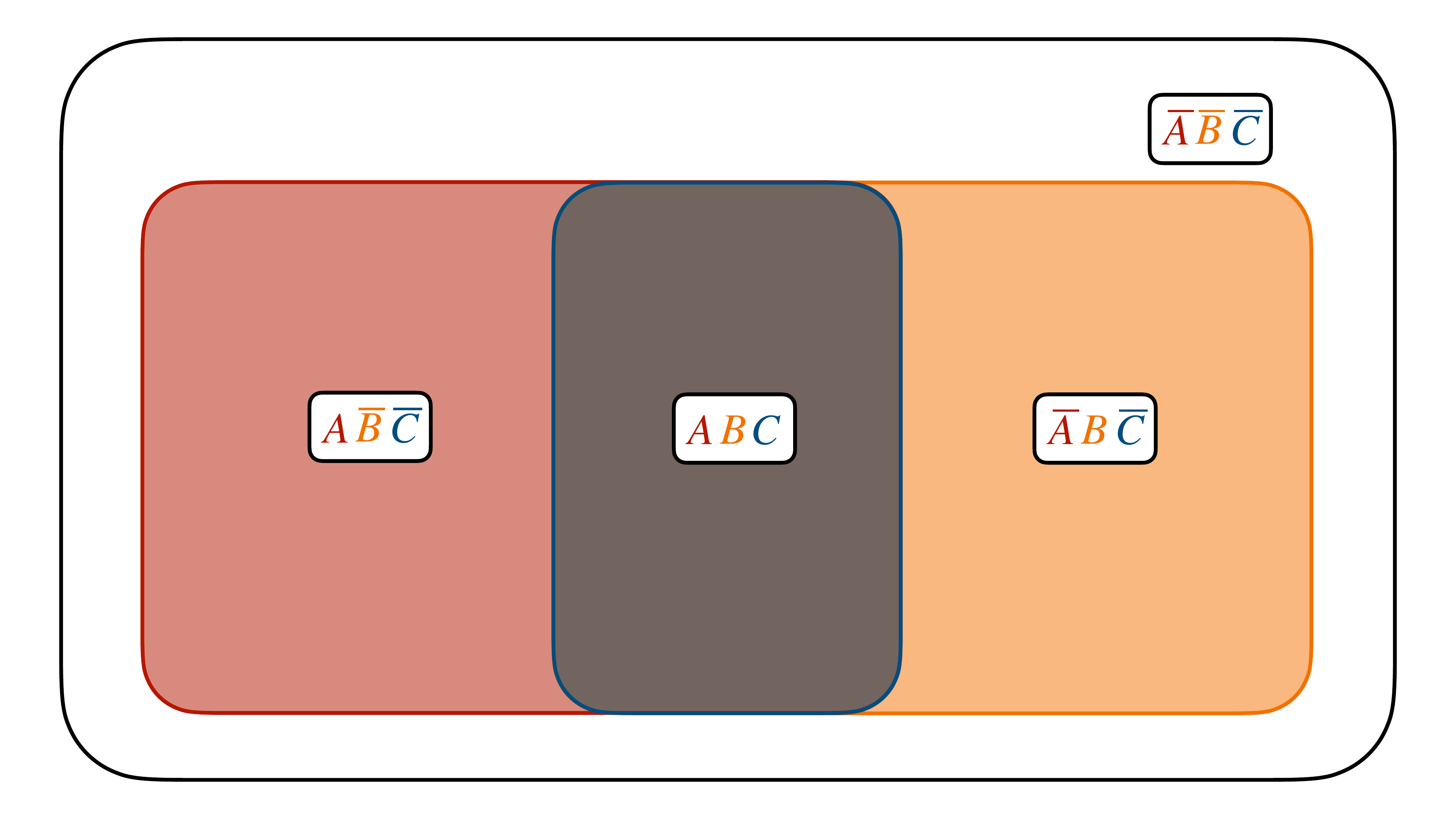}
	\end{subfigure}
    \begin{subfigure}[t]{0.5\textwidth}
        \caption{\textit{Category II-a} }
        \label{Fig:EulerDiagram3}
        \includegraphics[width=\textwidth]{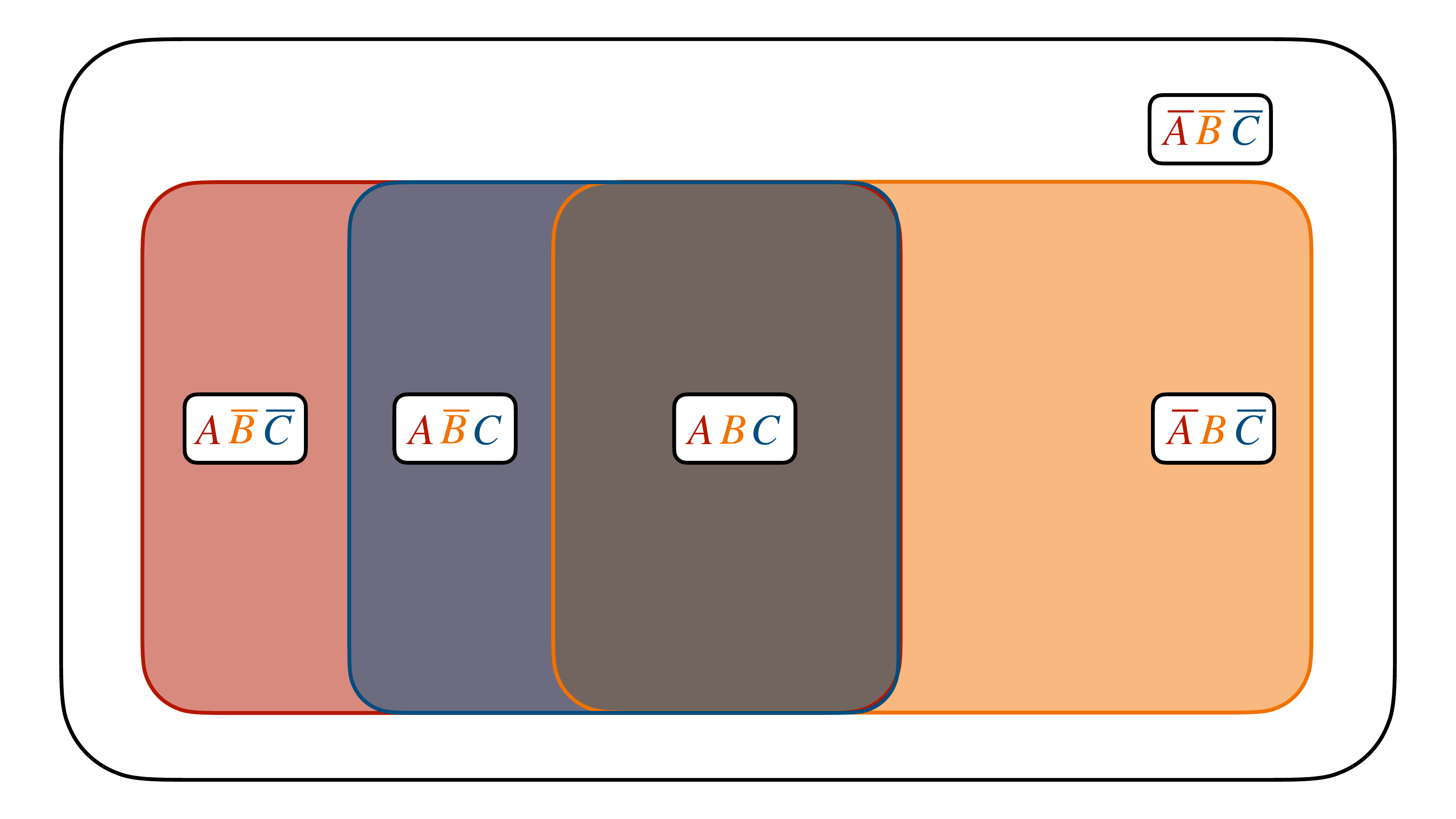}
	\end{subfigure}
	}
	\centerline{
    \begin{subfigure}[t]{0.5\textwidth}
        \vspace{0.5cm}
        \includegraphics[width=\textwidth]{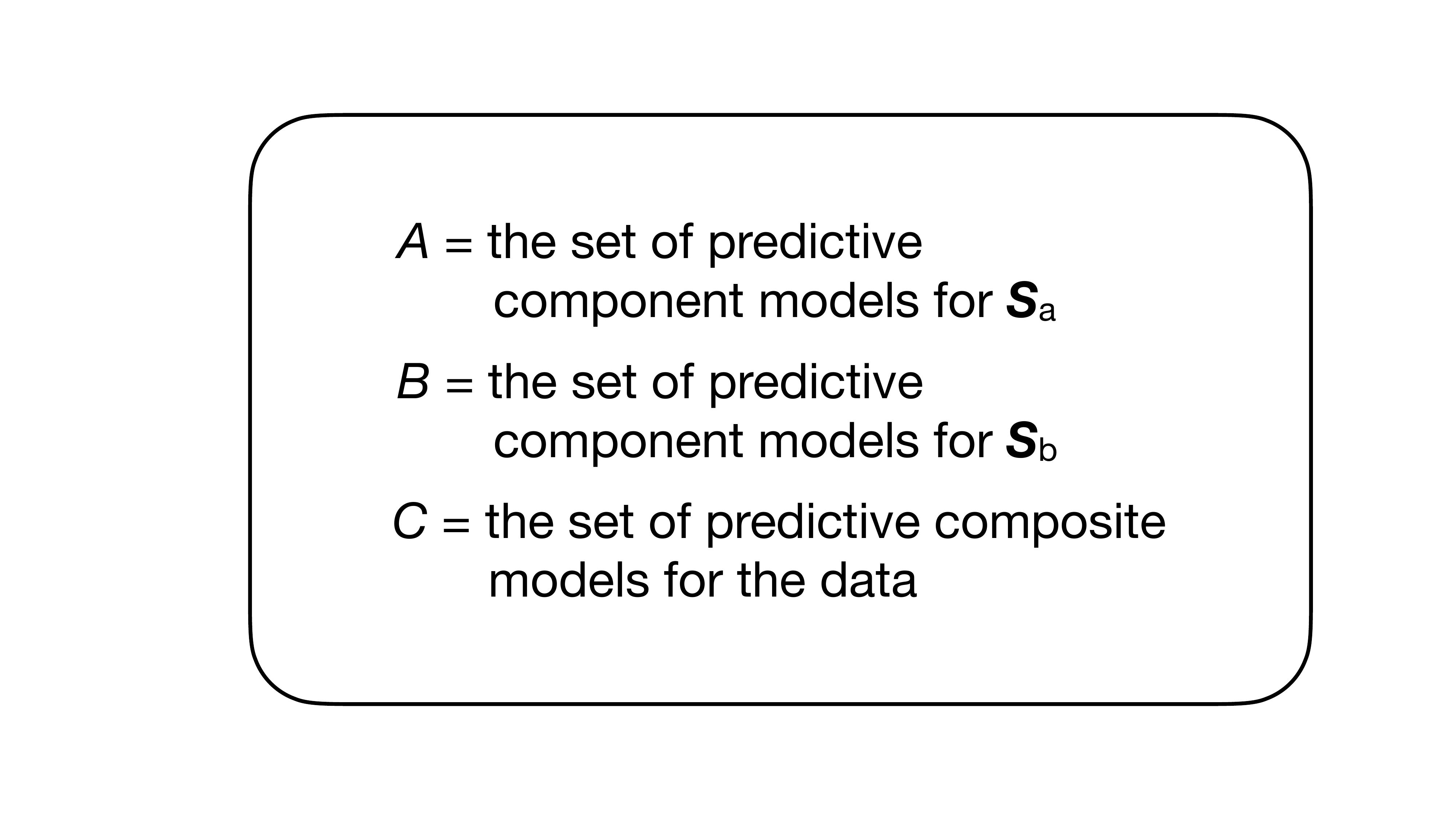}
    \end{subfigure}
	\begin{subfigure}[t]{0.5\textwidth}
        \caption{\textit{Category II-b}}
        \label{Fig:EulerDiagram2}
        \includegraphics[width=\textwidth]{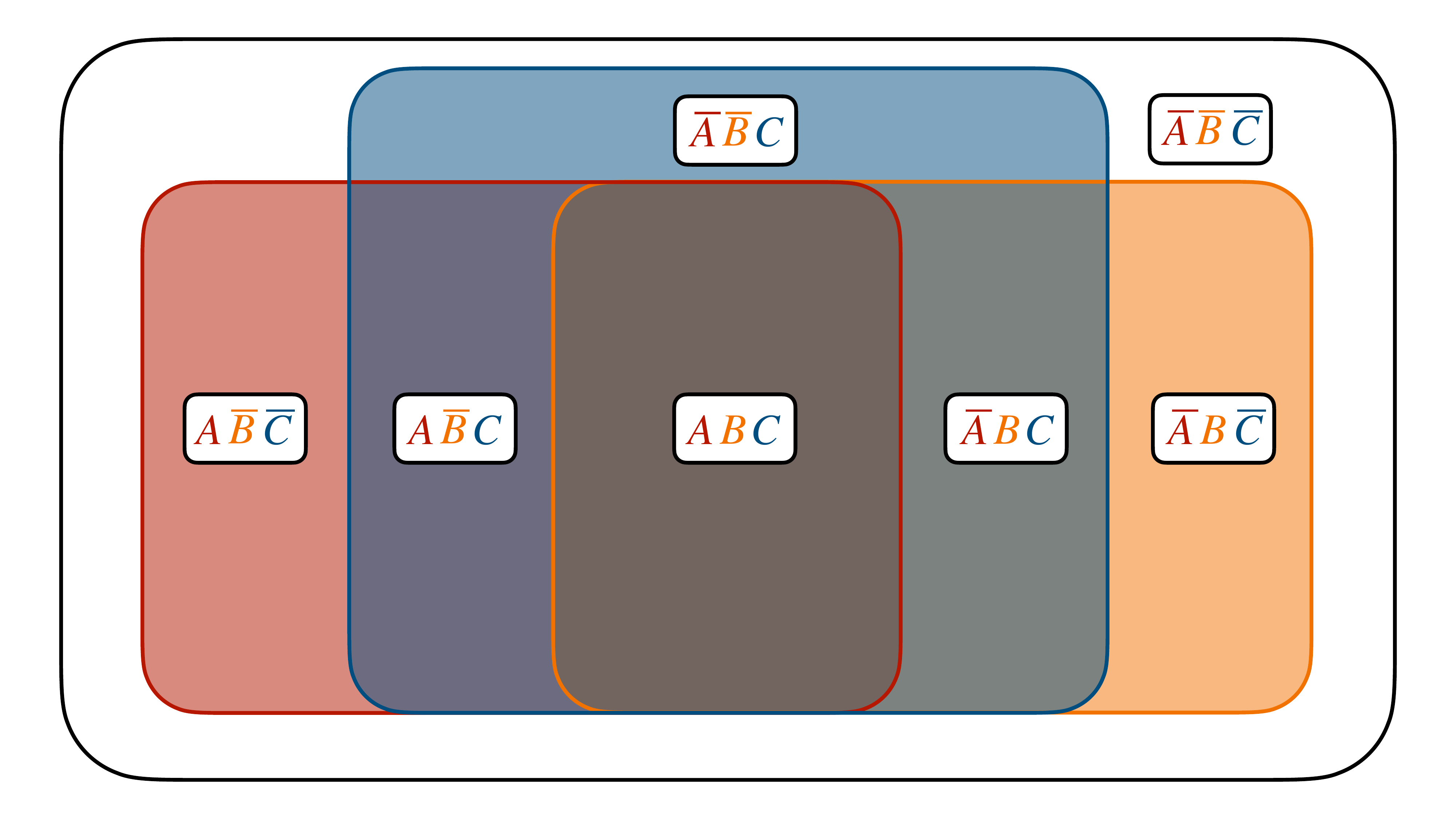}
	\end{subfigure}
	}
    \caption{
        Euler diagrams representing the logical relations between a set of competing composite models for a data set $\bm{D} = \bm{S}_\mathrm{a} + \bm{S}_\mathrm{b} + \bm{n}$ in \textit{category I} and \textit{II} model comparison scenarios. $A$, $B$ and $C$ represent the Euler sets of predictive and accurate models for $\bm{S}_\mathrm{a}$, $\bm{S}_\mathrm{b}$ and the data in aggregate, respectively. In each plot, `logical and' is represented by arithmetic multiplication (e.g. $ABC$ denotes the Euler set of models for which $A$, $B$ and $C$ are true.), and $\overline{A}$, $\overline{B}$ and $\overline{C}$ denote the logical complements to $A$, $B$ and $C$ (e.g. the Euler set of component models for $\bm{S}_\mathrm{a}$, $\bm{S}_\mathrm{b}$ with relatively low predictivity and the Euler set of composite models with relatively low predictivity, respectively).
        \textbf{Panel (a).}
        Euler diagram representing a \textit{`category I'} scenario in which the Euler set of predictive composite models for the data derives exclusively from the intersection of composite models containing predictive component models for $\bm{S}_\mathrm{a}$ and $\bm{S}_\mathrm{b}$.
        \textbf{Panel (b).} Euler diagram representing the \textit{category II-a} scenario in which the Euler set of predictive composite models for the data contains two subsets: the Euler set of predictive composite models with predictive component models for $\bm{S}_\mathrm{a}$ and $\bm{S}_\mathrm{b}$ that we are interested in ($ABC$) and the Euler set of predictive composite models with component models for $\bm{S}_\mathrm{a}$ (the SOI) that are predictive and for $\bm{S}_\mathrm{b}$ (nuisance-structure in the data) that have low predictivity ($A\overline{B}C$).
        \textbf{Panel (c).} Euler diagram representing the \textit{category II-b} scenario in which the Euler set of predictive composite models for the data contains four subsets: the Euler set of predictive composite models with predictive component models for $\bm{S}_\mathrm{a}$ and $\bm{S}_\mathrm{b}$ that we are interested in ($ABC$) and the Euler sets of predictive composite models with component models for $\bm{S}_\mathrm{a}$ or $\bm{S}_\mathrm{b}$ or both that have low predictivity ($\overline{A}BC$, $A\overline{B}C$ and $\overline{A}~\overline{B}C$, respectively).
    }
\label{Fig:EulerDiagrams}
\end{figure*}

\Cref{Fig:EulerDiagrams} shows Euler diagrams\footnote{An Euler diagram is a visual tool used to represent sets and their relationships. Unlike Venn diagrams, which show all possible set relations, Euler diagrams show only relevant relationships.} describing the relationships between the sets $A$, $\overline{A}$, $B$, $\overline{B}$, $C$, and $\overline{C}$ in three model comparison categories.

For clarity, and to distinguish them from $\bm{\mathcal{M}}_{\mathrm{a}}$, $\bm{\mathcal{M}}_{\mathrm{b}}$, $\bm{\mathcal{M}}_{\mathrm{c}}$, and other sets referred to in the remainder of the paper, we will refer to these sets, defined by the predictivity and accuracy conditions in \Cref{Eq:PredictivityCondition,Eq:AccuracyCondition}, as `Euler sets'.

The Euler diagram in \Cref{Fig:EulerDiagram1} illustrates the logical relationships between these Euler sets in a scenario where the component models must be elements of $A$ and $B$ for their composite to be an element of $C$ -- that is, when accurate and predictive composite models for the data exclusively consist of accurate and predictive component models for their signal counterparts.

In this case, all elements of $C$ are elements of $ABC$. Thus, the more predictive model will necessarily also have more accurate recovery of the signal components. Consequently, BFBMC provides a direct means of estimating the relative probability (weighted by model complexity and its ability to fit the data) that a model falls into the $ABC$ set (or equivalently, that it will yield unbiased parameter estimates). As such, \Cref{Fig:EulerDiagram1} illustrates the logical relations between Euler sets in a \textit{category I} model comparison.

\subsubsection{Category II Model Comparison}
\label{Sec:CategoryII}

\Cref{Fig:EulerDiagram2,Fig:EulerDiagram3} illustrate more complex scenarios in which composite models in $C$ are drawn from multiple subsets. In \Cref{Fig:EulerDiagram2}, the composite models in $C$ are drawn from four subsets: the set of accurate and predictive composite models with accurate and predictive component models for $\bm{S}_\mathrm{a}$ and $\bm{S}_\mathrm{b}$ ($ABC$), and the sets of accurate and predictive composite models with components for $\bm{S}_\mathrm{a}$, $\bm{S}_\mathrm{b}$, or both, that have low accurate and/or predictivity ($\overline{A}BC$, $A\overline{B}C$, and $\overline{A}~\overline{B}C$, respectively).

Models in $\overline{A}BC$ have components $\bm{M}_{j\mathrm{a}}$ that are low- accuracy and/or predictivity models for $\bm{S}_\mathrm{a}$. However, in the composite model $\bm{M}_{i\mathrm{c}}$, they are paired with components $\bm{M}_{k\mathrm{b}}$ that have sufficient flexibility to model both the underlying signal $\bm{S}_\mathrm{b}$ (if present) and the structure in $\bm{S}_\mathrm{a}$ that cannot be captured by $\bm{M}_{j\mathrm{a}}$. Similarly, $A\overline{B}C$ describes composite models with components $\bm{M}_{k\mathrm{b}}$ that are low- accuracy and/or predictivity models for $\bm{S}_\mathrm{b}$. In these composites, the $\bm{M}_{k\mathrm{b}}$ models are paired with components $\bm{M}_{j\mathrm{a}}$ that have sufficient flexibility to absorb any underlying signal $\bm{S}_\mathrm{a}$ (if present) and structure in $\bm{S}_\mathrm{b}$ that cannot be modelled by $\bm{M}_{k\mathrm{b}}$. Finally, $\overline{A}~\overline{B}C$ describes composite models in which both components $\bm{M}_{j\mathrm{a}}$ and $\bm{M}_{k\mathrm{b}}$ are low- accuracy and/or predictivity models for $\bm{S}_\mathrm{a}$ and $\bm{S}_\mathrm{b}$, respectively. However, $\bm{M}_{j\mathrm{a}}$ has sufficient flexibility to absorb structure in $\bm{S}_\mathrm{b}$ that cannot be modelled by $\bm{M}_{k\mathrm{b}}$ and vice versa.

\Cref{Fig:EulerDiagram3} illustrates a simplified model comparison scenario in which composite models in $C$ are drawn from two subsets: $ABC$ and $A\overline{B}C$.

Both \Cref{Fig:EulerDiagram2,Fig:EulerDiagram3} illustrate \textit{category II} model comparison problems.The logical relations between competing models for a data set simplify from \Cref{Fig:EulerDiagram2} to \Cref{Fig:EulerDiagram3} when $\overline{A}~BC$ and $\overline{A}~\overline{B}C$ are empty sets (i.e., have no elements). This occurs when the models fulfil one of the following two conditions:
\begin{itemize}
    \item \textbf{Condition 1:} There are no component models for $\bm{S}_\mathrm{b}$ that can absorb structure in $\bm{S}_\mathrm{a}$ that is impossible to model with $\bm{M}_{j\mathrm{a}}$. Thus, all composite models containing components from $\overline{A}$ are elements of $\overline{C}$.
    \item \textbf{Condition 2:} There exists a definitive model, $\bm{M}_{\mathrm{a}}$, such that there are no component models for $\bm{S}_\mathrm{a}$ in $\overline{A}$.
\end{itemize}

In this paper, we derive a framework for unbiased inference of $\bm{S}_\mathrm{a}$ from a data set when either of the above conditions is fulfilled, and the logical relations between the competing models for the data are as illustrated in \Cref{Fig:EulerDiagram3} (\textit{category II-a}). For concreteness, we assume that \textit{Condition 2} is fulfilled: we have a definitive model for $\bm{S}_\mathrm{a}$, $\bm{M}_{\mathrm{a}}$, which is an element of set $A$\footnote{Given a definitive model for $\bm{S}_{\mathrm{a}}$, it follows that $\bm{\mathcal{M}}_{\mathrm{a}} = \{\bm{M}_{\mathrm{a}}\}$. Correspondingly, there is a 1-to-1 mapping between the composite and nuisance component models under consideration. To reflect this, we label the nuisance models with subscript $i$ rather than $k$, with $\bm{M}_{i\mathrm{c}} = \bm{M}_{\mathrm{a}} + \bm{M}_{i\mathrm{b}}$.}. Thus, the set of models under consideration contains no elements in $\overline{A}$, and the composite models are elements of one of the following three sets: $ABC$, $A\overline{B}C$, or $A\overline{B}~\overline{C}$.

In practice, if one does not have a definitive model for $\bm{S}_\mathrm{a}$, the methodology described in this work can be used to derive a set of conditionally validated composite models. In this case, the validation of the composite models is contingent on the probability of $\bm{M}_{\mathrm{a}}$. If there are several candidate models for $\bm{S}_\mathrm{a}$, the validation methodology may be applied for each model, and the conclusions weighted by the relative probabilities assigned to each model of $\bm{S}_\mathrm{a}$. We leave to future work a more detailed examination of the generalization to \textit{category II-b} scenarios. Going forward, for brevity, we drop the \textit{‘a’} label and describe analyses and model comparisons categorised by the Euler diagram in \Cref{Fig:EulerDiagram3} as \textit{category II} data analyses and model comparisons, respectively.

To aid intuition for the model comparison classification introduced in this and the preceding section, the interested reader can find a discussion of how it applies to a concrete example in the context of 21-cm cosmology in \Cref{Sec:CategoryIandII21-cmCosmologyExample}.

\subsubsection{Transforming a Category II Model Comparison Problem into a Category I Model Comparison Problem}
\label{Sec:CategoryIItoCategoryI}

A \textit{category II} model comparison scenario is more likely when either or both of the component model sets contain flexible models with a priori poorly constrained parameters. In general, this increases the likelihood that $\bm{M}_{\mathrm{a}}$ can absorb residual structures arising from imperfect modelling of $\bm{S}_{\mathrm{b}}$ by $\bm{M}_{i\mathrm{b}}$. \textit{Category I} model comparison represents a special limiting case of \textit{category II} model comparison, where $A\overline{B}C$ is the empty set. Approaching this limit becomes more likely when one can reduce the volume of the model space under comparison using either implicit or explicit experimental\footnote{For example, through the analysis of alternate datasets that depend on the same underlying model.} or theory-based priors.

In a \textit{category II} scenario, BFBMC still allows the separation of models in set $C$ from those in its complement, $\overline{C}$. This disfavours composite models with components that fail to provide a consistent model of their signal counterparts \textit{and} cannot produce the observed data by any means. However, BFBMC does not enable one to distinguish between models of comparable complexity in the set $ABC$, for which one expects to recover unbiased estimates of the signal components, and those in the set $A\overline{B}C$, for which this is not the case. Therefore, using BFBMC alone limits one's ability to draw robust conclusions regarding which models will minimise bias in recovered parameter estimates.

However, if one can identify and exclude from the set of models under consideration those composite models in $A\overline{B}C$ a priori, then it is possible to transform a \textit{category II} model comparison problem into a \textit{category I} model comparison problem. We discuss a Bayesian-null-test-based methodology for achieving this transformation in the next section.

\subsection{BaNTER}
\label{Sec:BaNTER}

Let us define a validation data set in which the only signal component is $\bm{S}_\mathrm{b}$:
\begin{equation}
    \label{Eq:ValidationData}
    \bm{D}_\mathrm{v} = \bm{S}_\mathrm{b} + \bm{n} \ .
\end{equation}
Ideally one would use $\bm{S}_\mathrm{b}$-only observations for $\bm{D}_\mathrm{v}$; however, if obtaining such a data set is not experimentally feasible, realistic high-fidelity simulated $\bm{S}_\mathrm{b}$-only observations can provide a good substitute.

Two characteristic signatures of models in the set $A\overline{B}C$ are:
\begin{enumerate}
    \item the $\bm{M}_{i\mathrm{b}}$ component of $\bm{M}_{i\mathrm{c}}$ is a non-predictive model of the $\bm{S}_\mathrm{b}$ component of the data,
    \item the sum of statistically significant systematics resulting from inaccuracies of $\bm{M}_{i\mathrm{b}}$ as a model for $\bm{S}_\mathrm{b}$ can be absorbed by $\bm{M}_{\mathrm{a}}$ such that $\bm{M}_{i\mathrm{c}} = \bm{M}_{\mathrm{a}} + \bm{M}_{i\mathrm{b}}$ yields a predictive fit to the data in aggregate.
\end{enumerate}

Fundamentally, we are interested in assessing whether unbiased estimates of the SOI will be obtained in a fit of $\bm{M}_{i\mathrm{c}} = \bm{M}_{\mathrm{a}} + \bm{M}_{i\mathrm{b}}$ to $\bm{D}$. Thus, to validate a model $\bm{M}_{i\mathrm{b}}$ in $\bm{\mathcal{M}}_{\mathrm{b}}$, one must answer the question: is $\bm{M}_{i\mathrm{b}}$ sufficiently predictive of $\bm{S}_\mathrm{b}$ for a fit of $\bm{M}_{i\mathrm{c}} = \bm{M}_{\mathrm{a}} + \bm{M}_{i\mathrm{b}}$ to $\bm{D}$ to yield unbiased inferences of $\bm{S}_\mathrm{a}$?

If one can identify and exclude models in $\overline{B}$ a priori, the corresponding composite models in $A\overline{B}C$ will also be eliminated from the set of models under consideration for $\bm{D}$. Thus, initially one may assume that characteristic (i), alone, is sufficient for defining a model validation metric. However, in practice, this approach has limitations.

\Cref{Eq:BayesEqnForModels} can be used to calculate the posterior probabilities of models in $\bm{\mathcal{M}}_{\mathrm{b}}$ for $\bm{D}_\mathrm{v}$ or the posterior odds of the models can be determined using BFBMC (see \Cref{Sec:BayesianInference}). The former provides an absolute measure of the model predictivities and the latter a measure of their relative predictivities; however, neither directly answers whether $\bm{M}_{i\mathrm{b}}$ is \textit{sufficiently} accurate for our science case. Ultimately, this question is dependent on the models in $\bm{\mathcal{M}}_{\mathrm{a}}$ as well as those in $\bm{\mathcal{M}}_{\mathrm{b}}$, with larger imperfections in $\bm{M}_{i\mathrm{b}}$ acceptable if they are not describable with $\bm{M}_{\mathrm{a}}$, since in this case they cannot bias the SOI parameter inference. Thus, in this work, we take an alternate validation approach that addresses this question directly.

\subsubsection{Identifying and excluding models in $A\overline{B}C$}

Combining characteristics (i) and (ii), it follows that models in $\bm{\mathcal{M}}_{\mathrm{c}}$ that are elements of the set $A\overline{B}C$ are identifiable with a null-test between $\bm{M}_{i\mathrm{b}}$ and $\bm{M}_{i\mathrm{c}} = \bm{M}_{\mathrm{a}} + \bm{M}_{i\mathrm{b}}$ as models for $\bm{D}_\mathrm{v}$. Here, a preference for $\bm{M}_{i\mathrm{c}}$ over $\bm{M}_{i\mathrm{b}}$ as a model for $\bm{D}_\mathrm{v}$ suggests that $\bm{M}_{i\mathrm{b}}$ provides an incomplete description of the structure in $\bm{S}_\mathrm{a}$, \textit{and} that a spurious fit of $\bm{M}_{\mathrm{a}}$ absorbs residual structure in the validation data that $\bm{M}_{i\mathrm{b}}$ alone cannot fit. In a fit of $\bm{M}_{i\mathrm{c}}$ to $\bm{D}$, the same residual structure modelled by $\bm{M}_{\mathrm{a}}$ in the validation data will bias the SOI parameter inference. This signifies that $\bm{M}_{i\mathrm{b}}$ is not \textit{sufficiently} predictive for our science case.

Since $\bm{M}_{i\mathrm{b}}$ is a component of $\bm{M}_{i\mathrm{c}}$, the latter will always provide an equally good or better maximum likelihood fit to the data. Thus, we judge the relative preference for $\bm{M}_{i\mathrm{c}}$ over $\bm{M}_{i\mathrm{b}}$ by the Bayes factor between the models\footnote{
    Here, we use the Bayes factor as our validation metric due to its general applicability and simple interpretability. However, if simpler to calculate for the problem at hand, one might achieve a similar goal with alternate validation metrics including information criteria such as the Bayesian, Akaike or Deviance Information Criteria (e.g. \citealt{2008ConPh..49...71T}) or comparison of Kolmogorov-Smirnov test $p$-values deriving from the residuals of fits of $\bm{M}_{i\mathrm{b}}$ and $\bm{M}_{i\mathrm{c}}$ to $\bm{D}_\mathrm{v}$.
}. Let us define a log-Bayes-factor between a composite model $\bm{M}_{i\mathrm{c}} = \bm{M}_{i\mathrm{b}} + \bm{M}_{\mathrm{a}}$ and the component model for the nuisance structure, $\bm{M}_{i\mathrm{b}}$, as
\begin{equation}
    \label{Eq:lnBcb}
    \ln(\mathcal{B}_\mathrm{cb}^\mathrm{v}) = \ln(\mathcal{Z}_\mathrm{c}^\mathrm{v} / \mathcal{Z}_\mathrm{b}^\mathrm{v}) \ ,
\end{equation}
where $\mathcal{Z}_\mathrm{c}^\mathrm{v} = \mathcal{P}(\bm{D}_\mathrm{v} \vert \bm{M}_{i\mathrm{c}})$ and $\mathcal{Z}_\mathrm{b}^\mathrm{v} = \mathcal{P}(\bm{D}_\mathrm{v} \vert \bm{M}_{i\mathrm{b}})$ are the Bayesian evidence of $\bm{M}_{i\mathrm{c}}$ and $\bm{M}_{i\mathrm{b}}$, respectively\footnote{$\mathcal{Z}_\mathrm{b}^\mathrm{v}$ and $\mathcal{Z}_\mathrm{c}^\mathrm{v}$ can be estimated using nested sampling (see \Cref{Sec:ComputationalTechniques}). Additionally, for composite models where $\bm{M}_{i\mathrm{c}}$ reduces to $\bm{M}_{i\mathrm{b}}$ for a specific set of parameters of $\bm{M}_{\mathrm{a}}$ one can use the computationally efficient Savage-Dicke method (e.g. \citealt{SavageDicke}) to calculate $\ln(\mathcal{B}_\mathrm{cb}^\mathrm{v})$ directly from the prior and posterior distributions of $\bm{M}_{i\mathrm{c}}$. In this case, one negates the computational expense associated with deriving $\mathcal{Z}_\mathrm{b}^\mathrm{v}$ and $\mathcal{Z}_\mathrm{c}^\mathrm{v}$.}. When $\ln(\mathcal{B}_\mathrm{cb}^\mathrm{v}) < 0$, $\bm{M}_{i\mathrm{b}}$ is preferred over $\bm{M}_{i\mathrm{c}}$ for $\bm{D}_\mathrm{v}$, as expected when fitting $\bm{S}_\mathrm{b}$-only validation data.

Model classes with validation Bayes factors in the range $0 < \ln(\mathcal{B}_\mathrm{cb}^\mathrm{v}) \le 3$ have a weak-to-moderate preference for $\bm{M}_{i\mathrm{c}}$ over $\bm{M}_{i\mathrm{b}}$ (odds in favour of $\bm{M}_{i\mathrm{c}}$ over $\bm{M}_{i\mathrm{b}}$ of 20-to-1 or less; see \Cref{Tab:aPosterioriPreference}). This suggests that inaccuracies in the nuisance model are sufficient to bias estimates of SOI if present. However, under a Bayesian 21-cm detection criterion that requires strong evidence in favour of the composite model over the model for nuisance structure ($\ln(\mathcal{B}_\mathrm{cb}) \ge 3$; see \Cref{Sec:21cmSignalDetection}) this is below the level at which we would consider $\bm{S}_\mathrm{a}$ to have been spuriously detected.

$\ln(\mathcal{B}_\mathrm{cb}^\mathrm{v}) \ge 3$, corresponds to a strong preference for $\bm{M}_{i\mathrm{c}}$ over $\bm{M}_{i\mathrm{b}}$ in the validation data, with odds in favour of $\bm{M}_{i\mathrm{c}}$ over $\bm{M}_{i\mathrm{b}}$ of 20-to-1 or better (see \Cref{Tab:aPosterioriPreference}). This indicates that nuisance model inaccuracies are severe enough to significantly bias 21-cm signal estimates if present, or to lead to a spurious detection of the SOI if it is absent.

Here, we set a threshold above which we judge a nuisance model to have failed the null test of: $\ln(\mathcal{B}_\mathrm{cb}^\mathrm{v}) \ge \ln(\mathcal{B}_\mathrm{threshold}^\mathrm{v})$, with $\ln(\mathcal{B}_\mathrm{threshold}^\mathrm{v}) = 0$. In general, higher values of $\ln(\mathcal{B}_\mathrm{threshold}^\mathrm{v})$ correspond to less stringent validation criteria and vice versa. Ultimately, the value used for $\ln(\mathcal{B}_\mathrm{threshold}^\mathrm{v})$ should be informed by the quality of the validation data. We come back to this in \Cref{Sec:Discussion}.

\subsubsection{SOI model dependence}
\label{Sec:SOImodelDependence}

While here we have assumed that we have a definitive model for $\bm{S}_{\mathrm{a}}$ that is an element of set $A$ (see \Cref{Sec:CategoryII}), more generally, since our proposed null test will generically identify the subset of models where the systematics resulting from an inaccurate fit of $\bm{M}_{i\mathrm{b}}$ to $\bm{S}_\mathrm{b}$ can be accurately fit with $\bm{M}_{\mathrm{a}}$, we should expect its results to be dependent on the specific choice of model for $\bm{M}_{\mathrm{a}}$. The more flexible the $\bm{M}_{\mathrm{a}}$ component model, the more likely it is that it will be able to accurately describe systematics resulting from an inaccurate $\bm{M}_{i\mathrm{b}}$. However, the use of the Bayes-factor between models for the validation data in BaNTER validation balances this against the larger Occam penalty associated with the greater complexity of $\bm{M}_{\mathrm{a}}$ (see \Cref{Sec:BayesianInference}). Thus, $\bm{M}_{i\mathrm{b}}$ will only fail the null-test if $\bm{M}_{i\mathrm{c}}$ provides a sufficiently better fit to the data to justify its increased complexity.

\subsubsection{Null test results as model priors}
\label{Sec:NullTestResultsAsModelPriors}

If the BaNTER results indicate a statistically significant preference for $\bm{M}_{i\mathrm{c}}$ over $\bm{M}_{i\mathrm{b}}$, $\bm{M}_{i\mathrm{b}}$ and $\bm{M}_{i\mathrm{c}}$ are judged to have failed the null test. Given this information, one can update their assessment of the probability of these models. Here we consider models that fail the validation null-test to have negligible probability as models with which unbiased inferences of the SOI can be recovered in subsequent analysis of the data.

\subsubsection{Identifying and excluding models in $A\overline{B}~\overline{C}$}

In addition to its primary purpose of identifying models in $A\overline{B}C$, the BaNTER results will additionally identify models that produce a spurious detection of $\bm{S}_\mathrm{a}$ in the set $A\overline{B}~\overline{C}$\footnote{This is not guaranteed to identify all models in $A\overline{B}~\overline{C}$ since the logical possibility exists for systematics resulting from inaccuracies of $\bm{M}_{i\mathrm{b}}$ as models for $\bm{S}_\mathrm{b}$ to be of a form not modellable with $\bm{M}_{\mathrm{a}}$. Nevertheless, as such models will be elements of $\overline{C}$ they will still be identified by BFBMC and any inferences from them will be appropriately downweighted according to the degree to which they are disfavoured as models for the observational data, $\bm{D}$.}. Unlike models in $A\overline{B}C$, which we need validation data to identify, we expect models that are elements of $A\overline{B}~\overline{C}$ to independently be identified and disfavoured through BFBMC applied to the observational data. However, by disfavouring or eliminating all models that fail the null test, the additional disfavouring of models in $A\overline{B}~\overline{C}$ a priori can be treated as an ancillary benefit of the validation analysis.

\subsubsection{Using null test results and model accuracy to determine model comparison classification}
\label{Sec:ValidationFitResidualsTheory}

A composite model's failure of the validation null test identifies it as an element of either $A\overline{B}~\overline{C}$ or $A\overline{B}C$. One can obtain unbiased inferences regarding the SOI by downweighting models that fail the validation null-test irrespective of which of these two sets they are elements. Thus, determining which of the two sets a failed model is an element of is not an essential component of our proposed model-validated Bayesian inference workflow discussed in the next section. Nevertheless, confirmation that models can be correctly identified as elements of $A\overline{B}~\overline{C}$ and $A\overline{B}C$ using the validation analysis provides an additional cross-check of the methodology, and thus we briefly consider how it can be achieved.

One can predict whether a composite model that fails BaNTER validation is an element of $A\overline{B}~\overline{C}$ or $A\overline{B}C$ based on its ability to fit the validation data. If, in addition to failing the null test, the composite model does not meet the accuracy condition given in \Cref{Eq:AccuracyCondition}, this implies it is likely to be an element of $A\overline{B}~\overline{C}$. In contrast, if it fulfils the accuracy condition, albeit using a spurious fit of the SOI model, it is more likely to be an element of $A\overline{B}C$.

\subsection{Bayesian inference workflow}
\label{Sec:ModelValidatedBayesianInference}

\begin{figure*}
    \centerline{
        \includegraphics[width=\textwidth]{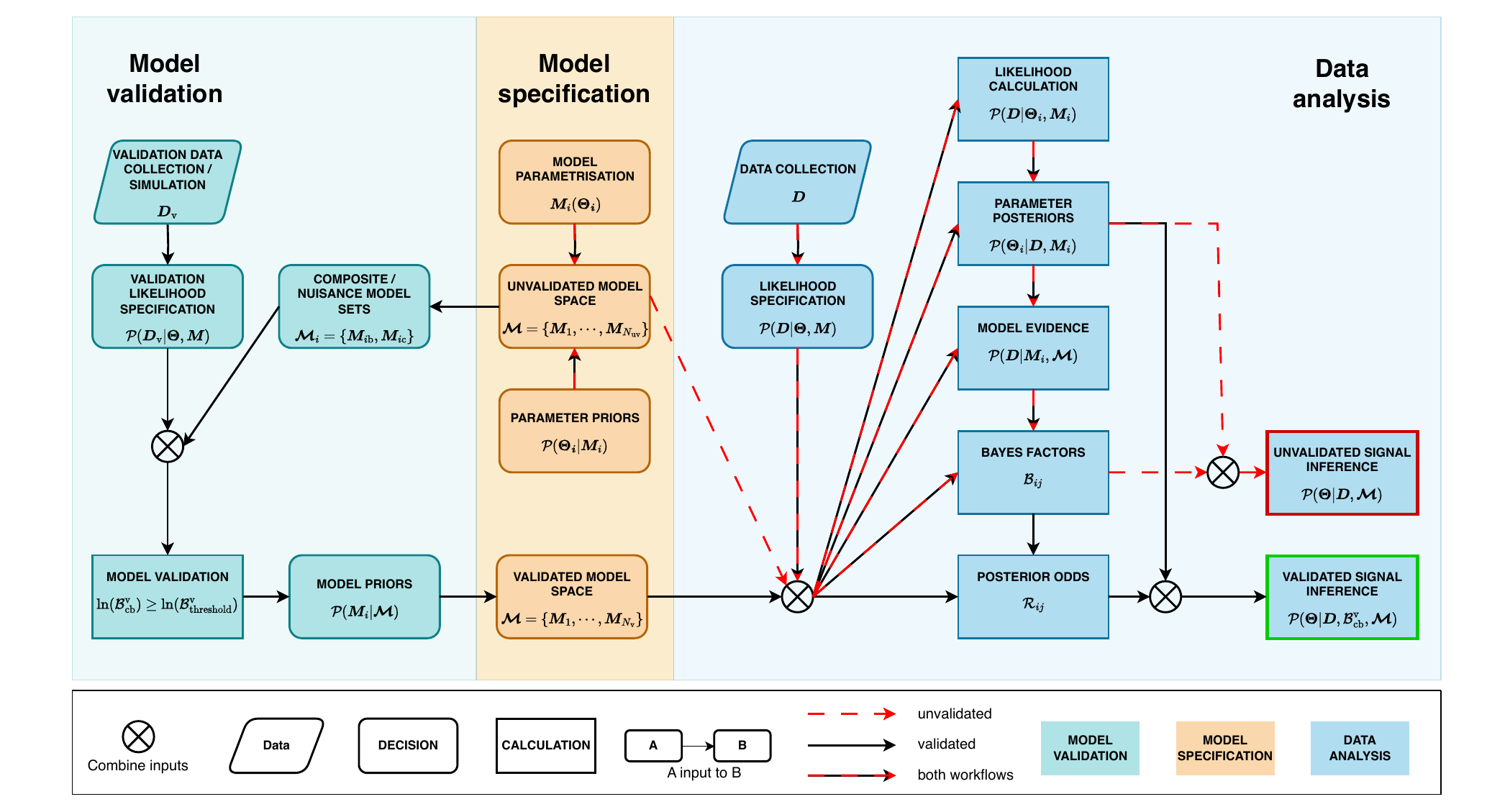}

        }
        \caption{
            Flowchart of a model-validated Bayesian inference workflow incorporating the BaNTER validation methodology introduced in this work. The workflow is divided into three main sections: model specification (orange), model validation (green) and data analysis (blue). Steps joined by dashed-red lines describe a \textit{Bayesian workflow with uninformative model priors} (see \Cref{Sec:BasicBayesianInference}). The model-validated Bayesian inference framework we propose in this work augments this workflow with BaNTER model validation. Steps joined by solid black lines describe our \textit{Model-validated Bayesian inference workflow} (see \Cref{Sec:ModelValidatedBayesianInferenceWorkflow}). Steps joined by solid black overlaid by dashed red lines are used in both workflows.
    }
\label{Fig:BayesianInferenceFlowChart}
\end{figure*}

\Cref{Fig:BayesianInferenceFlowChart} shows a flowchart of our BaNTER model validation framework and the way it interfaces with a typical Bayesian inference workflow. We divide the resulting model-validated Bayesian inference workflow into three main sections: model specification (orange), model validation (green) and data analysis (blue). The steps joined by dashed-red lines describe a \textit{typical Bayesian workflow with uninformative model priors}. The steps joined by solid black lines describe the \textit{model-validated Bayesian inference workflow} proposed in this work. We describe each below.

\subsubsection{Bayesian workflow with uninformative model priors}
\label{Sec:BasicBayesianInference}

In the model specification component of a typical Bayesian framework, one begins by specifying a set of models, $\bm{\mathcal{M}}$, that are considered as candidate models for the data. In \Cref{Fig:BayesianInferenceFlowChart} we label this as the `unvalidated model space'. For each model in this set one specifies the parameter priors, $\mathcal{P}(\sTheta_{i} \vert \bm{M}_{i})$ and the model parametrisation $\bm{M}_{i}(\sTheta_{i})$ that defines the mapping between $\sTheta_{i}$ and the modelled observables that will be compared with the data.

In the data analysis component, one collects the observational data, $\bm{D}$, and specifies a data likelihood, $\mathcal{P}(\sTheta\vert\bm{D},\bm{M})$ that reflects the noise covariance structure of the data. One then uses the unvalidated candidate models to fit the data according to the assumed likelihood. Conceptually, the analysis can then be divided into two tasks\footnote{In practice, if one uses a nested sampling algorithm to draw samples from the prior, as done in this work (see \Cref{Sec:ComputationalTechniques}), model evidence and parameter estimation are carried out simultaneously, with the latter being a by-product of the former.}: (i) parameter estimation, in which one draws samples from the parameter priors, calculates the corresponding data likelihood and posterior probability and uses the posterior samples to estimate the posterior density function of the model parameters; (ii) model comparison, in which one calculates the model evidence, $\mathcal{Z}_{i}$, and the Bayes factor between models, $\mathcal{B}_{ij}$ (see \Cref{Sec:BayesianInference} for details).

In the unvalidated model case, the Bayes factors define the relative probabilities of the models. If the weight of evidence in favour of one model over the others is sufficiently large, one may perform model selection. Alternatively and more generally, the model evidences can be used as weights for the parameter posteriors when deriving Bayesian model averaged parameter estimates and posterior predictive distributions for the signal components. In \Cref{Fig:BayesianInferenceFlowChart} we label these as unvalidated signal inference. If the unvalidated model space is well approximated by \textit{category I} model comparison, unbiased inferences can be obtained at this stage. If instead it represents a \textit{category II} model comparison problem, this procedure will yield biased inferences resulting from undue probability being assigned to biased inferences from models in $A\overline{B}C$.

\subsubsection{Model-validated Bayesian inference workflow}
\label{Sec:ModelValidatedBayesianInferenceWorkflow}

In the model-validated Bayesian inference workflow we begin by defining the unvalidated model space, as before. The BaNTER model validation component of the workflow is then responsible for identifying composite models in the unvalidated model space that are elements of the set $ABC$ versus $A\overline{B}C$ or $A\overline{B}~\overline{C}$, enabling one to separate the former from the latter and update the probability one assigns to the models in advance of analysing the observational data $\bm{D}$\footnote{Alternatively, one could achieve an equivalent validated signal inference by applying BaNTER validation just to the preferred set of models identified via BFBMC in the model comparison stage of the Bayesian workflow with uninformative model priors described in \Cref{Sec:BasicBayesianInference}. In this case, the initial model comparison is used to separate models in $ABC$ and $A\overline{B}C$ from those in $A\overline{B}~\overline{C}$ and BaNTER is used to distinguish between models in $ABC$ and $A\overline{B}C$.}. In this process, the models that fail the validation analysis are either,
\begin{enumerate}
    \item discarded from the set of models for the observational data under consideration, or,
    \item assigned a lower prior model probability than those which pass.
\end{enumerate}
In this work we treat the composite models that fail the validation null test as having negligible probability as models with which one may extract unbiased inferences of the SOI. As such, the above two options are equivalent. We use the second approach because it enables additional cross-checks regarding the Euler set that models which fail model validation derive from and, correspondingly, the category of the model comparison problem that is being undertaken. However, more generally, by eliminating the computational expense associated with deriving the Bayesian analysis data products from fits of the models that fail the validation analysis to the observational data, the first approach benefits from improved computational efficiency. Therefore, in future work, when applying the model-validated Bayesian inference workflow rather than demonstrating features of the model validation framework, the first approach may be preferable. We refer to the subset of models that pass the validation null test, which we consider credible models for deriving unbiased inferences regarding the SOI, as the `validated model space'.

The data analysis component of the model-validated Bayesian inference workflow proceeds in the same manner as the typical workflow with the following update. At the model comparison stage, one uses the parameter posteriors and posterior odds (rather than Bayes factors\footnote{The posterior odds are used implicitly if one has discarded models that fail the validation analysis from the set of models for the observational data under consideration. However, the practical procedure is to weight the model posteriors according to their Bayes factors (in the same manner as the typical workflow, but here applied to the validated model subset). In contrast, if models that fail the validation analysis are downweighted but not discarded, when calculating the posterior odds the reduced probability assigned to those models explicitly enters \Cref{Eq:BayesEqnForModels}, at this stage, via the model prior terms.}) to derive Bayesian model averaged parameter estimates and posterior predictive distributions for the signal components via \Cref{Eq:BayesEqnForModels}. This validated signal inference will yield unbiased estimates of the SOI in both \textit{category I} and \textit{II} model comparison problems\footnote{A caveat is that one should assign a probability to the results of the validation analysis in proportion to the quality of the validation data. Thus, if one lacks high-quality validation data this reduces the ability to rule out models that fail model validation with high confidence. We consider this issue further, and ways in which its impact can be mitigated, in \Cref{Sec:Discussion}.}.

\subsubsection{Model accuracy condition}
\label{Sec:ModelAccuracyCondition}

In both the unvalidated and validated Bayesian inference workflows discussed in \Cref{Sec:BasicBayesianInference,Sec:ModelValidatedBayesianInferenceWorkflow}, respectively, we assumed that $\bm{\mathcal{M}}_{\mathrm{c}}$ contains at least one model that is an element of $C$. If this is uncertain a priori, before proceeding with Bayesian model averaged parameter estimation we can additionally validate that the preferred models identified by model comparison meet the model accuracy condition given in \Cref{Eq:AccuracyCondition}. Assuming they do, having also passed the BaNTER null test, they can be identified as elements of $ABC$. In contrast, if they do not, they are judged to have failed model validation and considered insufficiently accurate for reliable inferences of the SOI. We assign negligible probability to these model, the same approach as with those that fail the Bayesian null test.

%%%%%%%%%%%%%%%%%%%%%%%%%%%%%%%%%%%%%%%%%%%%%%%%%%
\section{Application to global 21-cm cosmology}
\label{Sec:Results}
%%%%%%%%%%%%%%%%%%%%%%%%%%%%%%%%%%%%%%%%%%%%%%%%%%

Determining whether robust inferences from a set of models requires \textit{category I} or \textit{II} model comparison holds significant importance in the field of global 21-cm cosmology. The signal measured in a radiometric global 21-cm experiment can be written as the sum of two components: (A) a faint sky-averaged 21-cm cosmology signal (that we're interested in) and (B) intense foreground emission (a nuisance signal), both of which propagate through an instrumental transfer function. One would expect \textit{category II} model comparison to be necessary in global 21-cm cosmology if a subset of the foreground model components of a competing set of models cannot fully describe the foreground component of the data, but one or more of the composite models under consideration includes a 21-cm model that is capable of fitting the sum of systematics resulting from the inaccuracy of the foreground model and the 21-cm signal in the data (if present).

In this section we apply the BaNTER model validation methodology, described in \Cref{Sec:SIMVAMPTheory}, to an example model comparison problem motivated by global 21-cm cosmology data modelling. We consider three sets of data models ($\bm{\mathcal{M}}_{1}$, $\bm{\mathcal{M}}_{2}$ and $\bm{\mathcal{M}}_{3}$) that together illustrate salient features of the model validation methodology. We analyse the data using both the \textit{Bayesian workflow with uninformative model priors} and the \textit{model-validated Bayesian inference workflow} described in \Cref{Sec:ModelValidatedBayesianInference}. This enables us to demonstrate that only with the latter workflow can one:
\begin{itemize}
    \item  distinguish between sets of models requiring \textit{category I} and \textit{II} model comparison,
    \item identify models in $A\overline{B}C$ if present,
    \item ascertain that the composite models in the aforementioned three sets are elements of the $ABC$, $A\overline{B}C$ and $A\overline{B}~\overline{C}$, respectively,
    \item recover unbiased inferences of the cosmological 21-cm signal in the \textit{category II} model comparison problem.
\end{itemize}

%%%%%%%%%%%%%%%%%%%%%%%%%%%%%%%%%%%%%%%%%%%%%%%%%%
\subsection{Data models \& mock data}
\label{Sec:SimulatedDataAndDataModels}
%%%%%%%%%%%%%%%%%%%%%%%%%%%%%%%%%%%%%%%%%%%%%%%%%%

We consider a set of three competing composite 1D spectral models for a mock global 21-cm signal data set $\bm{D} = \bm{S}_\mathrm{a} + \bm{S}_\mathrm{b} + \bm{n}$, where $\bm{S}_\mathrm{a}$ is the 21-cm signal and $\bm{S}_\mathrm{b}$ is foreground emission\footnote{In practice, the instrument and ionosphere impart multiplicative effects on the astrophysical signals such that the data is more accurately represented as $\bm{D} = f(\bm{S}_\mathrm{a} + \bm{S}_\mathrm{b}) + \bm{n}$. However, these effects are small under ideal conditions regarding ionospheric conditions and instrumental calibration and here, for simplicity, we neglect them and treat the data as the sum of independent components.}:
\begin{itemize}
    \item Model 1: $\bm{M}_{1\mathrm{c}} = \bm{M}_\mathrm{a} + \bm{M}_\mathrm{1b}$
    \item Model 2: $\bm{M}_{2\mathrm{c}} = \bm{M}_\mathrm{a} + \bm{M}_\mathrm{2b}$
    \item Model 3: $\bm{M}_{3\mathrm{c}} = \bm{M}_\mathrm{a} + \bm{M}_\mathrm{3b}$
\end{itemize}
Each composite model has a 21-cm signal component with a functional form matching the true generative model for the 21-cm signal in the mock data ($\bm{M}_\mathrm{a}$) and one of three foreground model components ($\bm{M}_{1\mathrm{b}}$, $\bm{M}_{2\mathrm{b}}$ and $\bm{M}_{3\mathrm{b}}$). Going forward, we use $\bm{\mathcal{M}}_{i} = \{\bm{M}_{i\mathrm{c}}, \bm{M}_{i\mathrm{b}}\}$ to denote the $i$th composite/nuisance model set. When analysing the data, we assume that $\bm{S}_\mathrm{b}$ is known to be present in the data, a priori; however, the presence, or not, of $\bm{S}_\mathrm{a}$ is unknown. The full set of models under consideration for the data is given by $\bm{\mathcal{M}} = \{\bm{M}_{1\mathrm{c}}, \bm{M}_{1\mathrm{b}}, \bm{M}_{2\mathrm{c}}, \bm{M}_{2\mathrm{b}}, \bm{M}_{3\mathrm{c}}, \bm{M}_{3\mathrm{b}}\}$. Prior to model validation or analysis of the data we treat each of the models in $\bm{\mathcal{M}}$ as equally probable.

\subsubsection{21-cm signal model}

We model the 21-cm signal as a flattened-Gaussian absorption trough with a functional form matching the model parametrisation used in \citet[hereafter B18]{2018Natur.555...67B}:
\begin{equation}
\label{Eq:FlattenedGaussian}
\bm{M}_\mathrm{a} = \bm{T}_{21} = -A\left(\frac{1-\e^{-\tau \e^{\bm{B}_{21}}}}{1-\e^{-\tau}}\right) \ ,
\end{equation}
with,
\begin{equation}
\label{Eq:FlattenedGaussianB}
\bm{B}_{21} = \frac{4(\bm{\nu}-\nu_0)^2}{w^2}\log\left[-\frac{1}{\tau}\log\left(\frac{1+\e^{-\tau}}{2}\right)\right] \ .
\end{equation}
Here, $A$, $\nu_0$, $w$ and $\tau$ describe the amplitude, central frequency, width and flattening of the absorption trough, respectively, and $\bm{\nu}$ is a vector of central frequencies of the channels over the observing band of the data.

\subsubsection{Foreground models}

As the basis of our foreground description we use the class of models defined by the flexible closed-form parametrisation derived for beam-factor chromaticity corrected radio spectrometer data
\footnote{In Sims et al. (in preparation), we test a set of competing models
drawn from several model classes that have been applied to 21-cm signal estimation in the literature. We find that, assuming high-fidelity modelling of the instrument, models of the form given in \Cref{Eq:BFCCdataModel} are most predictive of beam-factor chromaticity corrected radio spectrometer data.
}
in S23:
\begin{multline}
    \label{Eq:BFCCdataModel}
    \bm{T}_\mathrm{BFCC}^\mathrm{Fg}(\sTheta_\mathrm{BFCC})
    = \Bigg[\bar{T}_\mathrm{m_{0}}\left(\frac{\bm{\nu}}{\nu_\mathrm{c}}\right)^{-\beta_{0}}(1 + \sum_{\alpha=1}^{N_\mathrm{pert}} p_{\alpha}\ln\left(\frac{\bm{\nu}}{\nu_\mathrm{c}}\right)^{\alpha}) \\
    + \frac{(1-\left(\frac{\bm{\nu}}{\nu_\mathrm{c}}\right)^{-\beta_{0}}) T_{\gamma}}{\bar{\bm{B}}_\mathrm{factor}(\bm{\nu})}\Bigg] \e^{-\tau_\mathrm{ion}(\bm{\nu})} + \frac{T_{\mathrm{e}}}{\bar{\bm{B}}_\mathrm{factor}(\bm{\nu})}(1-\e^{-\tau_\mathrm{ion}(\bm{\nu})}) \ .
\end{multline}
Here, all operations on vector quantities are element-wise, and we parametrise the effective ionospheric optical depth as,
\begin{equation}
    \label{Eq:tau_ion}
    \tau_\mathrm{ion} = \tau_0(\nu/\nu_\mathrm{c})^{-2} \ ,
\end{equation}
where $\nu_\mathrm{c} = 75~\mathrm{MHz}$ is a reference frequency. The free parameters in \Cref{Eq:BFCCdataModel} are $\sTheta_\mathrm{BFCC} = [\bar{T}_\mathrm{m_{0}}, \beta_0, T_{\mathrm{e}}, \tau_0, p_1, \cdots, p_{N_\mathrm{pert}}]$. We describe these parameters and provide an overview of the individual terms in the equation in \Cref{Sec:ForegroundModelDescription}. For a detailed description of the physical and instrumental effects described by \Cref{Eq:BFCCdataModel} we refer the reader to S23. This, however, is not a prerequisite for understanding the features of Bayesian model validation and comparison that are the focus of this paper.

We define our three foreground component models using the following variants of the \Cref{Eq:BFCCdataModel} model class:
\begin{itemize}
    \item $\bm{M}_{1\mathrm{b}}$ has $N_\mathrm{f} = 7$ fitted foreground parameters, including: the amplitude and spectral index of the foreground emission ($\bar{T}_\mathrm{m_{0}}$ and $\beta_0$), the temperature of ionospheric electrons and the effective optical depth of the ionosphere (at reference frequency $\nu_\mathrm{c}$) through which the foreground emission travels ($T_{\mathrm{e}}$ and $\tau_0$, respectively) and $N_\mathrm{pert} = 3$ terms for describing the amplitude of instrumentally coupled spectral fluctuations about the sky-averaged spectrum of the foreground brightness temperature field. Hereafter, for brevity, we refer to the $N_\mathrm{pert}$ model components as instrument-foreground coupling terms. As will be discussed in \Cref{Sec:MockData}, we use $\bm{M}_{1\mathrm{b}}$ as a generative model for $\bm{S}_\mathrm{a}$ when constructing the mock data.
    \item $\bm{M}_{2\mathrm{b}}$ has $N_\mathrm{f} = 5$ free parameters, matching those in model $\bm{M}_{1\mathrm{b}}$, but with only $N_\mathrm{pert} = 1$ instrument-foreground coupling term. Thus, this model assumes that structure in the data from higher than 1st order instrument-foreground coupling is negligible. If this assumption does not hold, meaning $\bm{M}_{2\mathrm{b}}$ provides a poor description of $\bm{S}_{\mathrm{b}}$, but the model $\bm{M}_{2\mathrm{c}}$ can fit the data accurately by biasing the foreground and the 21-cm signal parameters, the models in $\bm{\mathcal{M}}_{2}$ will fail the null test.
    \item $\bm{M}_{3\mathrm{b}}$ has $N_\mathrm{f} = 4$ free parameters, including the amplitude and spectral index of the foreground emission and $N_\mathrm{pert} = 2$ instrument-foreground coupling terms. This model assumes that structures in the data described by the 3rd order instrument-foreground coupling term and ionospheric effects are negligible. As with $\bm{M}_{2\mathrm{b}}$, if these assumptions do not hold, meaning $\bm{M}_{3\mathrm{b}}$ provides a poor description of $\bm{S}_{\mathrm{b}}$, but $\bm{M}_{3\mathrm{c}}$ can better fit the data by biasing the other foreground parameters and the 21-cm signal parameters, then the models in $\bm{\mathcal{M}}_{3}$ will fail the null test.
\end{itemize}

We note that for continuity, foreshadowing the results, the model labelling has been chosen such that the results with the composite models incorporating these components, $\bm{M}_{1\mathrm{c}}$, $\bm{M}_{2\mathrm{c}}$ and $\bm{M}_{3\mathrm{c}}$, respectively, align qualitatively with the properties of the corresponding numbered models discussed in the introduction.

\subsubsection{Model summary and parameter priors}

The structural properties of the full set of models under consideration for the data are summarised in \Cref{Tab:DataModelsSummary}. When fitting these models to the mock data we impose priors on the parameters of the component models that are summarised in \Cref{Tab:DataModelPriors2}. These priors have a combination of physical and experimental motivations (see S23 for details).
\begin{table*}
    \caption{
        A summary of the structural properties of the models in $\bm{\mathcal{M}} = \{\bm{M}_{1\mathrm{c}}, \bm{M}_{1\mathrm{b}}, \bm{M}_{2\mathrm{c}}, \bm{M}_{2\mathrm{b}}, \bm{M}_{3\mathrm{c}}, \bm{M}_{3\mathrm{b}}\}$.
    }
    \centerline{
    \begin{tabular}{l l l l l}
    \hline
    Model & Number of parameters & Instrument-foreground coupling & Includes ionospheric model & Includes 21-cm signal model     \\
    & & terms ($N_\mathrm{pert}$) & &      \\
    \hline
    $\bm{M}_{1\mathrm{c}}$ & $11$     & $3$ & Y & Y \\
    $\bm{M}_{2\mathrm{c}}$ & $9$     & $1$ & Y & Y \\
    $\bm{M}_{3\mathrm{c}}$ & $8$     & $2$ & N & Y \\
    $\bm{M}_{1\mathrm{b}}$ & $7$     & $3$ & Y & N \\
    $\bm{M}_{2\mathrm{b}}$ & $5$     & $1$ & Y & N \\
    $\bm{M}_{3\mathrm{b}}$ & $4$     & $2$ & N & N \\
    \hline
    \end{tabular}
    }
\label{Tab:DataModelsSummary}
\end{table*}

\begin{table}
    \caption{
        Priors on the parameters of the 21-cm signal and foreground model components. The relation between $\tau_0$ and $\tau_\mathrm{ion}$ is given in \Cref{Eq:tau_ion}.
    }
    \centerline{
    \begin{tabular}{l l l l }
    \hline
    Model component & Model & Parameter & Prior     \\
    \hline
    21-cm signal &  $\bm{M}_\mathrm{a}$ & $A$     & $U(0,1)~\mathrm{K}$ \\
    &  $\bm{M}_\mathrm{a}$ & $\nu_0$ & $U(55,95)~\mathrm{MHz}$ \\
    &  $\bm{M}_\mathrm{a}$ & $w$     & $U(5,30)~\mathrm{MHz}$ \\
    &  $\bm{M}_\mathrm{a}$ & $\tau$  & $U(0,20)$  \\
    \hline
    Foreground & $\bm{M}_{1\mathrm{b}}$, $\bm{M}_{2\mathrm{b}}$, $\bm{M}_{3\mathrm{b}}$ & $\bar{T}_\mathrm{m_{0}}$ & $U(1000,6000)~\mathrm{K}$ \\
    & $\bm{M}_{1\mathrm{b}}$, $\bm{M}_{2\mathrm{b}}$, $\bm{M}_{3\mathrm{b}}$ & $\beta_{0}$ & $U(2.0,3.0)$ \\
    & $\bm{M}_{1\mathrm{b}}$, $\bm{M}_{2\mathrm{b}}$ & $T_{\mathrm{e}}$ & $U(100,800)~\mathrm{K}$ \\
    & $\bm{M}_{1\mathrm{b}}$, $\bm{M}_{2\mathrm{b}}$ & $\tau_0$ & $U(0.005,0.025)$ \\
    & $\bm{M}_{1\mathrm{b}}$, $\bm{M}_{2\mathrm{b}}$, $\bm{M}_{3\mathrm{b}}$ & $p_{0}$ & $U(-0.1,0.1)$ \\
    & $\bm{M}_{1\mathrm{b}}$, $\bm{M}_{3\mathrm{b}}$ & $p_{1}$ & $U(-0.1,0.1)$ \\
    & $\bm{M}_{1\mathrm{b}}$ & $p_{2}$ & $U(-0.1,0.1)$ \\
    \hline
    \end{tabular}
    }
\label{Tab:DataModelPriors2}
\end{table}

\subsubsection{Mock data}
\label{Sec:MockData}

We generate mock data, $\bm{D}$, of the form given in \Cref{Eq:Data}, using $\bm{M}_\mathrm{a}$ and $\bm{M}_{1\mathrm{b}}$ as generative models for $\bm{S}_\mathrm{a}$ and $\bm{S}_\mathrm{b}$ such that $\bm{M}_{1\mathrm{c}}$ (for a specific set of model parameters) is the true generative model of the data.

\begin{figure*}
\centerline{
	\begin{subfigure}[t]{0.5\textwidth}
        \caption{}
        \label{Fig:T21}
        \includegraphics[width=\textwidth]{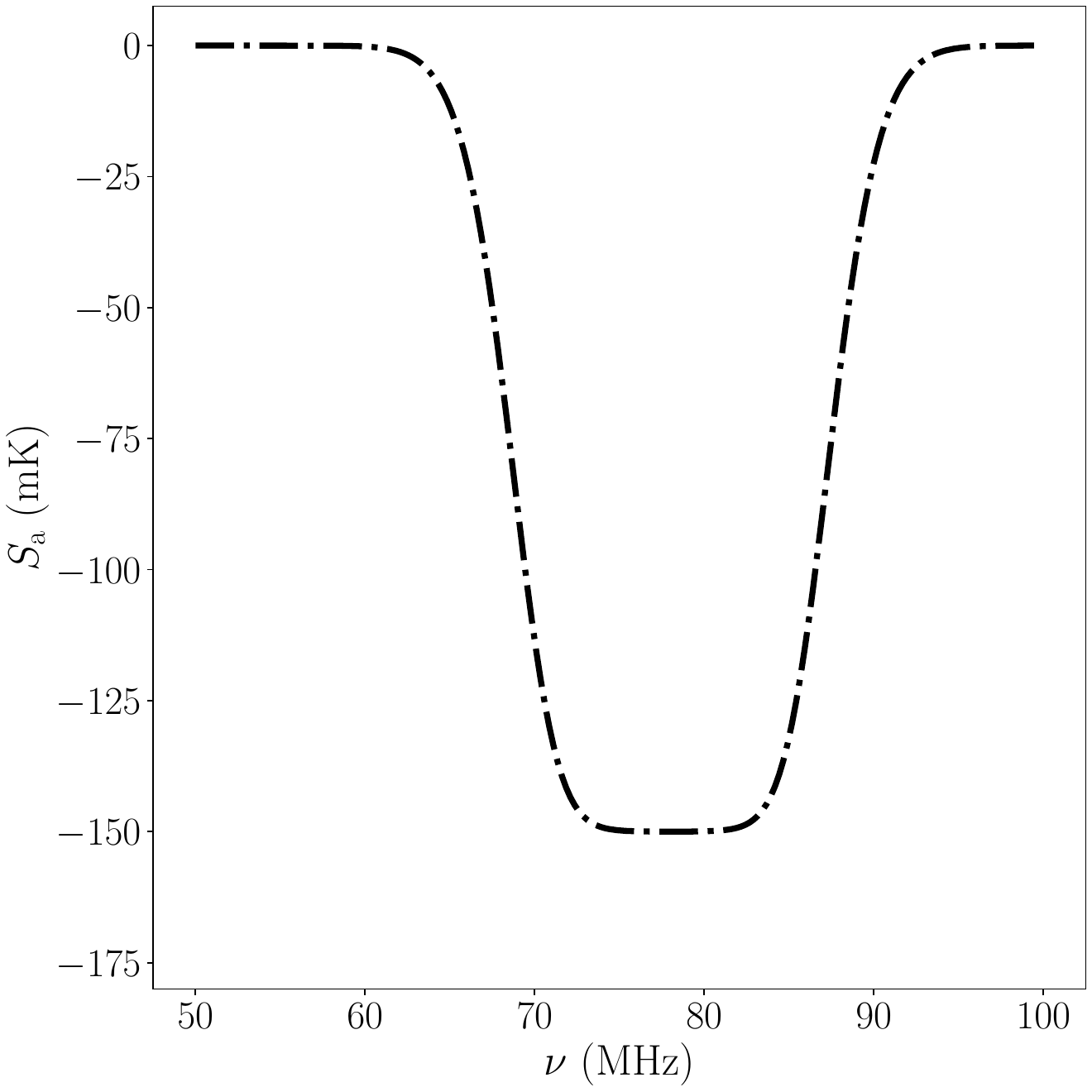}
	\end{subfigure}
	\begin{subfigure}[t]{0.5\textwidth}
        \caption{}
        \label{Fig:Tcorrected}
        \includegraphics[width=\textwidth]{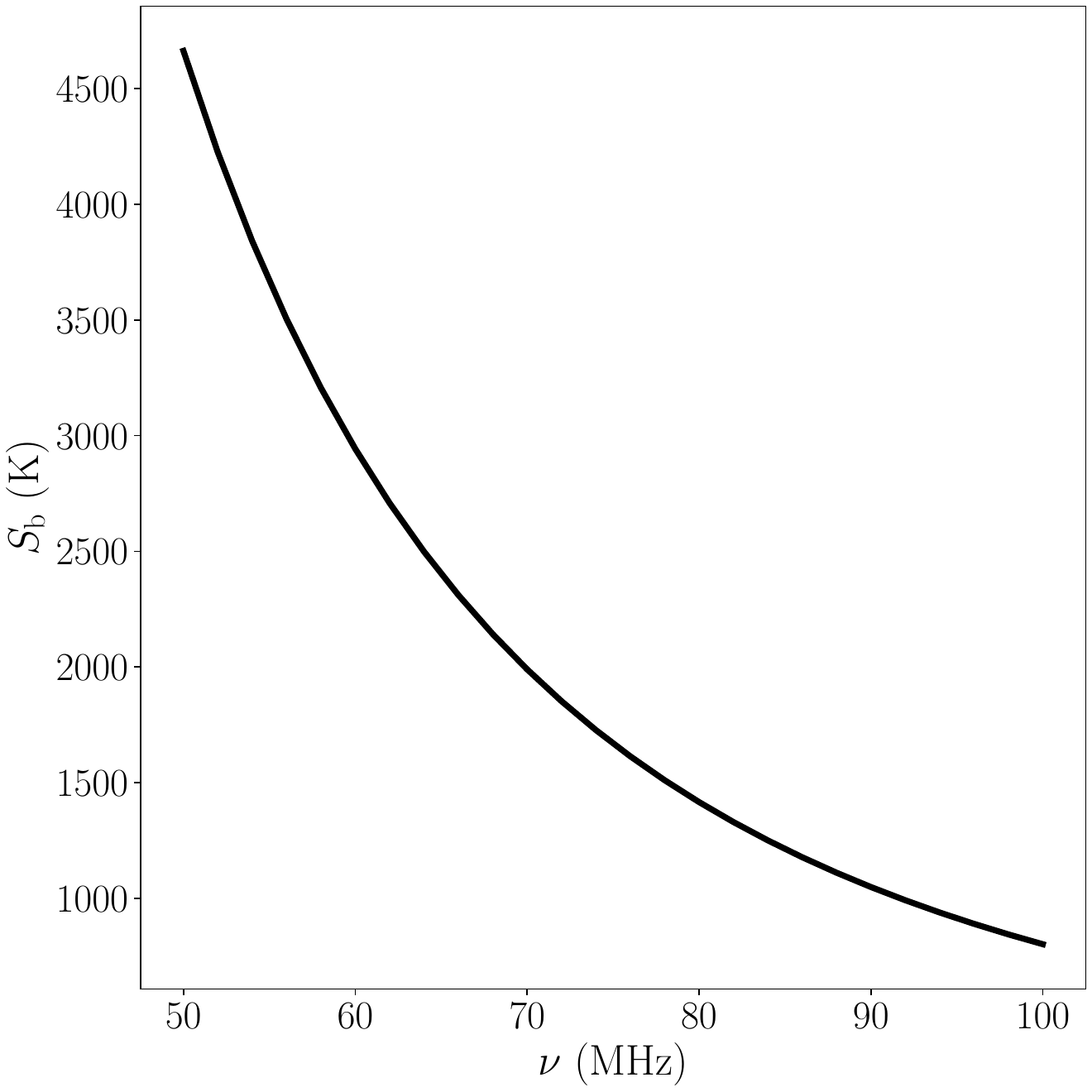}
	\end{subfigure}
	}
\caption{
    The signal components of the mock data.
    \textbf{Panel (a).} Signal component $\bm{S}_\mathrm{a}$ - a flattened-Gaussian absorption trough with an amplitude of $150~\mathrm{mK}$ and central frequency of $78~\mathrm{MHz}$.
    \textbf{Panel (b).} Signal component $\bm{S}_\mathrm{b}$ - radio foreground emission in beam-factor chromaticity corrected spectrometer data.
    Note the difference in units on the vertical axes of the two Panels. The amplitude of $\bm{S}_\mathrm{b}$ at $78~\mathrm{MHz}$ is $\sim 1500~\mathrm{K}$, which is 4 orders of magnitude brighter than the peak amplitude of $\bm{S}_\mathrm{a}$ at this frequency. As a result, spectral structure in $\bm{S}_\mathrm{b}$ unmodelled at the level of 1 part in $10^4$ by $\bm{M}_{i\mathrm{b}}$ is sufficient to leave systematic structure comparable in amplitude to $\bm{S}_\mathrm{a}$.
    }
\label{Fig:MockData}
\end{figure*}

We generate signals (A) and (B) as $\bm{S}_\mathrm{a} = \bm{M}_\mathrm{a}(\sTheta_\mathrm{a})$ and $\bm{S}_\mathrm{b} = \bm{M}_{1\mathrm{b}}(\sTheta_\mathrm{b})$, respectively. The values of the 21-cm parameter vector, $\sTheta_\mathrm{a} = [A, \nu_0, w, \tau]$, are listed in the top panel of \Cref{Tab:DataGenerationParameters}. They are selected to be consistent with the 21-cm signal estimates in B18, except for the amplitude of the signal, which we set to a value of $150~\mathrm{mK}$. The values of the foreground parameter vector, $\sTheta_\mathrm{b} = [\bar{T}_\mathrm{m_{0}}, \beta_0, p_1, p_2, p_3, T_\mathrm{e}, \tau_0]$ are listed in the middle panel of \Cref{Tab:DataGenerationParameters}. They are derived from the maximum likelihood fit of $\bm{M}_{1\mathrm{b}}$ to simulations of foreground emission as observed by the EDGES experiment (see S23). $\bm{S}_\mathrm{a}$ and $\bm{S}_\mathrm{b}$ are illustrated in \Cref{Fig:T21,Fig:Tcorrected}, respectively\footnote{We note that this definition of $\bm{S}_\mathrm{a}$ makes \Cref{Eq:FlattenedGaussian} a definitive model of the SOI for the purposes of fitting the mock data in this work (see \Cref{Sec:CategoryII}, \textit{condition 2}).
}.

To generate the mock data we add to the signal components zero-mean uncorrelated Gaussian random noise drawn from the distribution $N(\sTheta_\mathrm{n})$. The values of the noise parameter vector, $\sTheta_\mathrm{n} = [\mu_\mathrm{n}, \sigma_\mathrm{n}]$ are listed in the middle panel of \Cref{Tab:DataGenerationParameters}, and are chosen to produce a noise realisation with properties comparable to those estimated for the publicly available EDGES-low data from B18 (e.g. \citealt{2019ApJ...880...26S}).

\begin{table*}
    \caption{
        True values of the 21-cm signal parameters used to generate $\bm{S}_\mathrm{a}$, the foreground parameters used to generate $\bm{S}_\mathrm{b}$, and noise parameters used to generate $\bm{n}$. The generative model (GM) of the signal components is noted in the component descriptions.
        }
    \centerline{
    \begin{tabular}{l l l l l}
    \hline
    Data component & Component description & Parameter & Value & Parameter description     \\
    \hline
    $\bm{S}_\mathrm{a}$ & 21-cm signal & $A$ & $150~\mathrm{mK}$ & Absorption depth \\
                        & [GM: \Cref{Eq:FlattenedGaussian}] & $\nu_0$ & $78~\mathrm{MHz}$ & Central frequency \\
                        & & $w$     & $19~\mathrm{MHz}$ & Width \\
                        & & $\tau$  & $8$ & Flattening parameter \\
    \hline
    $\bm{S}_\mathrm{b}$ & Foreground emission & $\bar{T}_\mathrm{m_{0}}$ & $1627.46~\mathrm{K}$  & Foreground brightness temperature at $\nu = \nu_\mathrm{c}$ \\
    & [GM: \Cref{Eq:BFCCdataModel}] &     $\beta_{0}$                  & $2.488$ & Mean temperature spectral index  \\
    & &     $p_{1}$                      & $-0.080$ & 1st-order spectral fluctuations fractional amplitude \\
    & &     $p_{2}$                      & $-0.002$ & 2nd-order spectral fluctuations fractional amplitude \\
    & &     $p_{3}$                      & $0.013$ & 3rd-order spectral fluctuations fractional amplitude \\
    & &     $T_{\mathrm{e}}$             & $117.31~\mathrm{K}$ & Ionospheric electron temperature \\
    & &     $\tau_0$                     & $0.009$ & Ionospheric optical depth at $\nu=\nu_\mathrm{c}$ \\
    \hline
    $\bm{n}$ & Noise & $\mu_\mathrm{n}$ & $0~\mathrm{mK}$  & Noise mean \\
    & [GM: $N(\mu_\mathrm{n}, \sigma_\mathrm{n})$] &     $\sigma_\mathrm{n}$                  & $20~\mathrm{mK}$ & Noise standard deviation  \\
    \hline
    \end{tabular}
    }
\label{Tab:DataGenerationParameters}
\end{table*}

\subsubsection{Including the true generative model in the set of models under consideration}

In practice, all statistical models that are fit to observational data are approximate (e.g. \citealt{Box1976}); thus, having one of the models under consideration being the true generative model is an idealisation. However,  this test case has a number of benefits over a comparison involving only approximate models:
\begin{enumerate}
    \item it illustrates that \textit{category II} model comparison scenarios can arise in this limit and thus that they are not a by-product of comparing competing composite models for the data in which all models in the set under consideration are approximate,
    \item with this test-case, the preferred model for the data that one aims to identify with an ideal model comparison framework is evident a priori, simplifying the evaluation of the efficacy of the model validation framework and subsequent posterior-odds-based comparison of the models under consideration.
\end{enumerate}
Furthermore, a situation in which all models in the set under consideration are approximate can be expected to converge on that considered here, in the limit that at least a subset models for the data accurately describe it at a given signal-to-noise level. The test-case is thus also illustrative of this common situation\footnote{In Sims et al. (in preparation) we demonstrate that this expectation is borne out, in the context of fitting models to simulated global 21-cm signal data, when all models of the data under consideration are approximate.}.

\subsubsection{Validation data}
\label{Sec:ValidationData}

We construct validation data, $\bm{D}_\mathrm{v}$, of the form given in \Cref{Eq:ValidationData} that contains only the nuisance signal $\bm{S}_\mathrm{b}$. This is generated in the same manner as the mock data except, in this case, omitting $\bm{S}_\mathrm{a}$ from the data vector.

\subsubsection{Data Likelihood}
\label{Sec:DataLikelihood}

We include uncorrelated Gaussian white noise in our simulated data. We correspondingly model its covariance matrix, $\mathbfss{N}$, as diagonal, with elements given by:
\begin{equation}
    N_{ij} = \left< n_in_j^*\right> = \delta_{ij}\sigma^{2} \ .
\end{equation}
Here, $\left< .. \right>$ represents the expectation value and $\sigma = 20~\mathrm{mK}$.

Given this noise covariance model and defining a vectorised model for the signal, $\bm{m}$, constructed from a vector of parameters $\bm{\Theta}$ and a residuals vector between the data and model given by $\bm{r} = \bm{D} - \bm{m(\Theta)}$ we can specify a Gaussian likelihood for the simulated data as,
\begin{equation}
\label{Eq:GaussianLikelihood}
\mathcal{L}(\bm{\Theta}) = \frac{1}{\sqrt{(2\pi)^{N_{\mathrm{chan}}}\mathrm{det}(\mathbfss{N})}} \exp\left[-\frac{1}{2}\bm{r(\Theta)}^T\mathbfss{N}^{-1}\bm{r(\Theta)}\right] \ .
\end{equation}
We make use of this likelihood when fitting the validation and mock data in \Cref{Sec:SIMVAMPApplication,Sec:BFBMCvsPOBMC}, respectively.

\subsection{Signal detection}
\label{Sec:21cmSignalDetection}

Bayesian model comparison provides a means by which the detectability of $\bm{S}_\mathrm{a}$ can be assessed. Let us assume that we have a predictive composite model for the data of the form $\bm{M}_{i\mathrm{c}} = \bm{M}_\mathrm{a} + \bm{M}_{i\mathrm{b}}$, where $\bm{M}_{i\mathrm{c}}$ is an element of the set $ABC$. For a data set known to contain $\bm{S}_\mathrm{b}$, we can ascertain the evidence in favour of a detection of $\bm{S}_\mathrm{a}$ using BFBMC of $\bm{M}_{i\mathrm{c}}$ and $\bm{M}_{i\mathrm{b}}$ as models for the data. Given that $\bm{M}_{i\mathrm{b}}$ is a subcomponent of $\bm{M}_{i\mathrm{c}}$, a fit of $\bm{M}_{i\mathrm{c}}$ to the data will necessarily have a likelihood equal or higher than that of $\bm{M}_{i\mathrm{b}}$. However, by using the difference in Bayesian evidence of the models as the comparison metric, BFBMC of $\bm{M}_{i\mathrm{c}}$ and $\bm{M}_{i\mathrm{b}}$ will yield a preference for $\bm{M}_{i\mathrm{c}}$ over $\bm{M}_{i\mathrm{b}}$ only when $\bm{M}_{i\mathrm{c}}$ provides a fit that is improved sufficiently to counterbalance the Occam penalty associated with its additional complexity relative to $\bm{M}_{i\mathrm{b}}$.

We define the threshold for a detection of $\bm{S}_\mathrm{a}$, as $\ln(\mathcal{B}_\mathrm{cb}) \ge 3$, where $\ln(\mathcal{B}_\mathrm{cb})$ is as defined in \Cref{Eq:lnBcb} except, in this case, for the observational rather than validation data. This corresponds to odds in favour of $\bm{M}_{i\mathrm{c}}$ over $\bm{M}_{i\mathrm{b}}$ of 20-to-1 or better (see \Cref{Tab:aPosterioriPreference}) and is consistent with guidelines established by \cite{KandR} for strong evidence in favour of one model relative to another\footnote{In principle, one could also adopt a simulation-based inference approach for this task. For example, one could assess signal detection using a neural binary classifier (e.g. \citealt{2024arXiv240314618S}), trained on simulated examples of detections and non-detections. This approach offers flexibility and the ability to optimize the detection metric for signals of the sort expected in the data; however, it also requires additional computational resources and careful validation to ensure robustness. We leave this as a direction for future exploration.\label{Foot:SBI}}.

In the general case where it is unknown whether $\bm{M}{i\mathrm{c}}$ is an element of $ABC$, a preference for $\bm{M}{i\mathrm{c}}$ over $\bm{M}{i\mathrm{b}}$ indicates that $\bm{S}\mathrm{a}$ is either detected in the data or that a spurious detection has occurred due to the presence of systematics\footnote{We use the term 'systematics' in a general sense to refer to any residual structure in the data resulting from imperfect modelling of $\bm{S}\mathrm{b}$ by $\bm{M}{i\mathrm{b}}$. This can include unmodelled instrumental, environmental, or astrophysical effects, or a combination of these factors.}. Model validation is crucial for distinguishing between these two possibilities.

\subsection{Computational techniques}
\label{Sec:ComputationalTechniques}

In \Cref{Sec:SIMVAMPApplication,Sec:BFBMCvsPOBMC}, we estimate model evidences and sample from the posteriors of model parameters, given the data, using nested sampling as implemented by the \textsc{PolyChord} algorithm \citep{2015MNRAS.453.4384H, 2015MNRAS.450L..61H}. Given samples from the posterior distribution of the parameters, $\mathcal{P}(\sTheta\vert\bm{D},\bm{M})$, we can estimate the posterior predictive density (posterior PD), $\mathcal{P}(y\vert\sTheta,\nu,\bm{D},\bm{M})$, for a function $y = f(\sTheta,\nu)$. This is achieved by calculating the set of samples that correspond to $\mathcal{P}(y\vert \sTheta,\nu,\bm{D},\bm{M})$. We generate contour plots of prior and posterior PDs using the \textsc{fgivenx} software package (\citealt{2018JOSS....3..849H}).

In our analysis, the foreground and composite models are nested. The Savage-Dickey ratio provides a computationally efficient method for calculating the Bayes factor between nested models using the prior and posterior distributions of the most complex model, $\bm{M}_{1\mathrm{c}}$ (e.g., \citealt{SavageDicke}). In this approach, one calculates the ratio of the posterior and prior densities evaluated at zero of the components that the complex model has in addition to the nested model. However, for the purposes of confirming model accuracy and BaNTER model classification predictions, we require not only the Bayes factors but also the parameter posteriors of the individual models. Using nested sampling, we obtain both the parameter posteriors and Bayesian evidences; consequently, we do not explicitly employ the Savage-Dickey ratio approach in this work. Nevertheless, we note that it can be an efficient method for performing BaNTER validation and BFBMC when one's sole objective is to use these tools for validated Bayesian model comparison, without the additional objective of demonstrating the efficacy of the framework.

\subsection{BaNTER model validation results}
\label{Sec:SIMVAMPApplication}

\begin{table}
    \caption{
        Validation null test results. Bayes factors between $\bm{M}_{i\mathrm{c}}$ and $\bm{M}_{i\mathrm{b}}$, as models for the validation data, where $i$ runs over the 3 model classes defined in \Cref{Sec:SimulatedDataAndDataModels}. A positive $\ln(\mathcal{B}_\mathrm{cb}^\mathrm{v})$ means that $\bm{M}_{i\mathrm{c}}$ is preferred over $\bm{M}_{i\mathrm{b}}$ as a model of $\bm{D}_\mathrm{v}$. The reverse is true when $\ln(\mathcal{B}_\mathrm{cb}^\mathrm{v})$ is negative.
    }
    \centerline{
        \begin{tabular}{l l l}
            \hline
            Model class & $\ln(\mathcal{B}_\mathrm{cb}^\mathrm{v})$ & Comment \\
            \hline
            $\bm{\mathcal{M}}_{1}$ & -3.2 & Null test passed \\
            $\bm{\mathcal{M}}_{2}$ & 184.6 & Spurious detection - null test failed \\
            $\bm{\mathcal{M}}_{3}$ & 2997.2 & Spurious detection - null test failed \\
            \hline
    \end{tabular}
}
\label{Tab:lnBdetection1}
\end{table}

\subsubsection{Null test}
\label{Sec:NullTestResults}

\Cref{Tab:lnBdetection1} lists values of $\ln(\mathcal{B}_\mathrm{cb}^\mathrm{v})$ -- the Bayes factors between $\bm{M}_{i\mathrm{c}}$ and $\bm{M}_{i\mathrm{b}}$ as models for the validation data,  $\bm{D}_\mathrm{v}$ -- calculated for the 3 model classes defined in \Cref{Sec:SimulatedDataAndDataModels}. For positive $\ln(\mathcal{B}_\mathrm{cb}^\mathrm{v})$, $\bm{M}_{i\mathrm{c}}$ is preferred over $\bm{M}_{i\mathrm{b}}$. Under the assumption that $\bm{M}_{i\mathrm{c}}$ is an accurate model for $\bm{D}_\mathrm{v}$ this would constitute evidence for a detection of $\bm{S}_\mathrm{a}$; however, since $\bm{S}_\mathrm{b}$ is the only signal component that $\bm{D}_\mathrm{v}$ contains, any detection of $\bm{S}_\mathrm{a}$ in the validation data is necessarily spurious.

For $\bm{\mathcal{M}}_{1}$, $\ln(\mathcal{B}_\mathrm{cb}^\mathrm{v}) = -3.2$, which corresponds to a strong preference (see \Cref{Tab:aPosterioriPreference}) for $\bm{M}_{1\mathrm{b}}$, relative to $\bm{M}_{1\mathrm{c}}$, as a model for the validation data. As such, $\bm{\mathcal{M}}_{1}$ passes the model validation test. In contrast, $\bm{M}_{i\mathrm{c}}$ is decisively preferred over $\bm{M}_{i\mathrm{b}}$ for both $\bm{\mathcal{M}}_{2}$ and $\bm{\mathcal{M}}_{3}$ ($\ln(\mathcal{B}_\mathrm{cb}^\mathrm{v}) = 184.6$ and $2997.2$, respectively). Correspondingly, $\bm{\mathcal{M}}_{2}$ and $\bm{\mathcal{M}}_{3}$ fail the validation null test, and we can identify $\bm{M}_{2\mathrm{c}}$ and $\bm{M}_{3\mathrm{c}}$ as elements of either $A\overline{B}C$ or $A\overline{B}~\overline{C}$.

\begin{figure*}
    \centerline{
        \begin{subfigure}[t]{0.33\textwidth}
            \caption{\Large{$\bm{\mathcal{M}}_{1}$, PV}}
            \label{Fig:ValidationM1}
            \includegraphics[width=\textwidth]{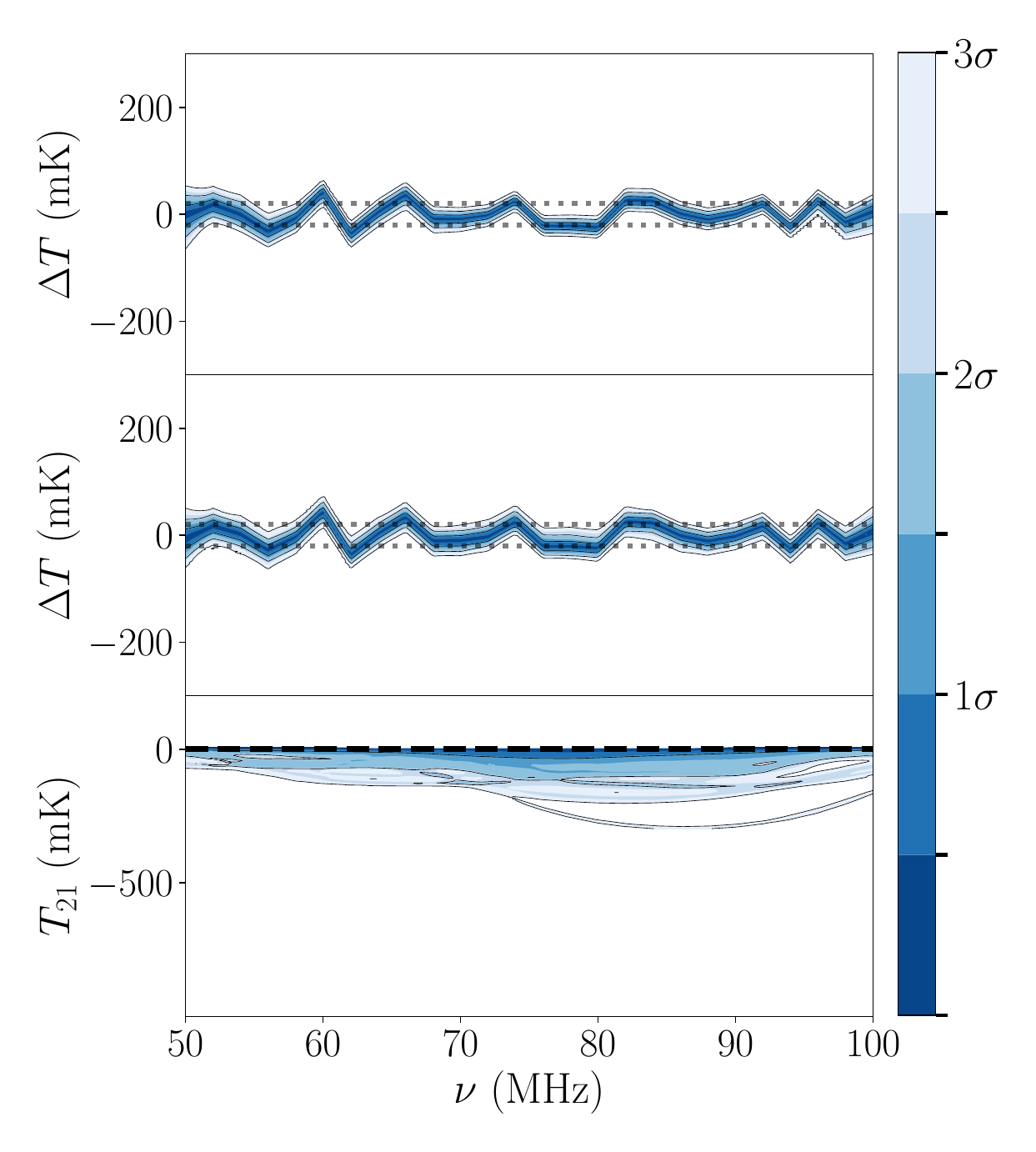}
        \end{subfigure}
        \begin{subfigure}[t]{0.33\textwidth}
            \caption{\Large{$\bm{\mathcal{M}}_{2}$, FV}}
            \label{Fig:ValidationM2}
            \includegraphics[width=\textwidth]{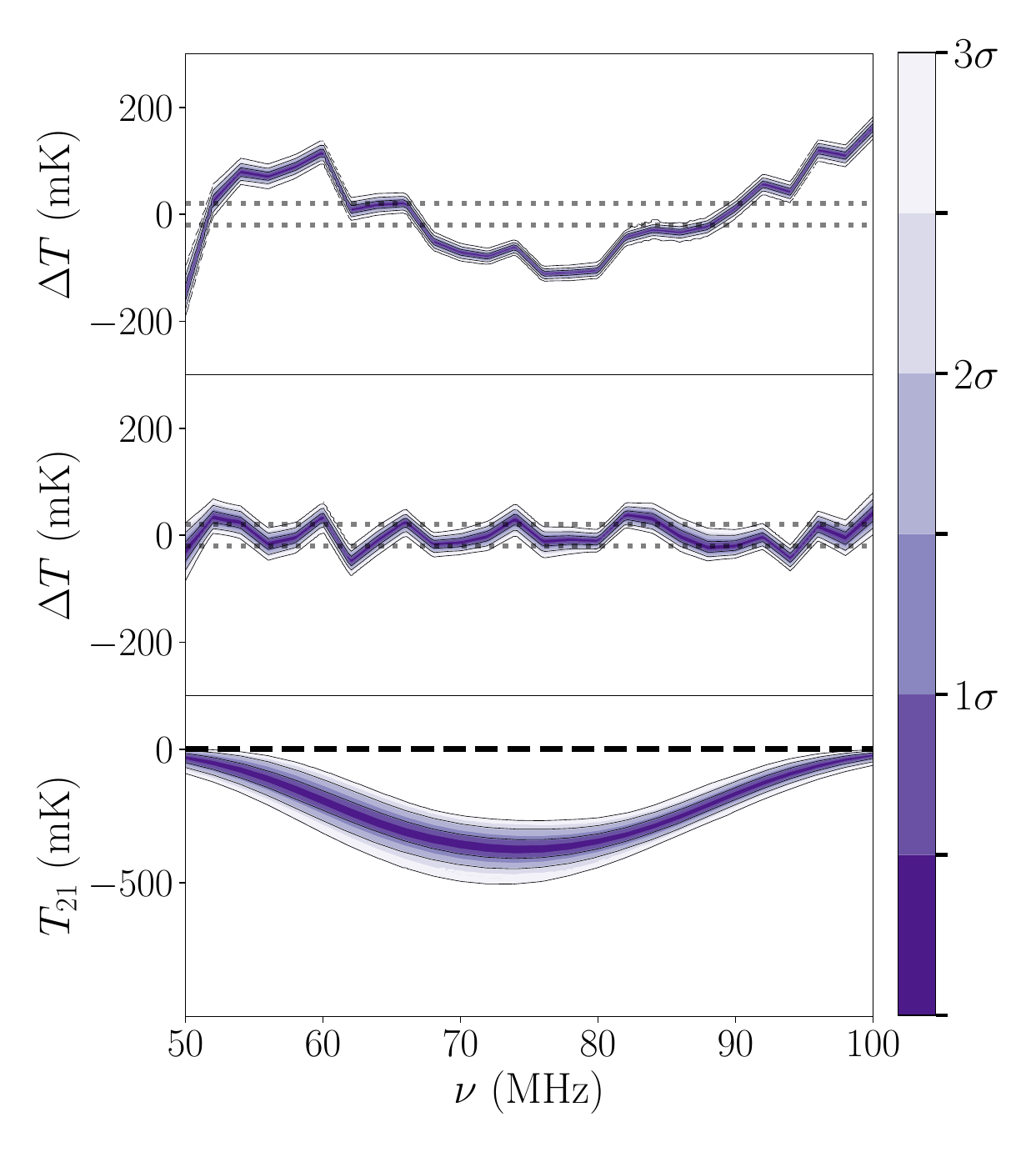}
        \end{subfigure}
        \begin{subfigure}[t]{0.33\textwidth}
            \caption{\Large{$\bm{\mathcal{M}}_{3}$, FV}}
            \label{Fig:ValidationM3}
            \includegraphics[width=\textwidth]{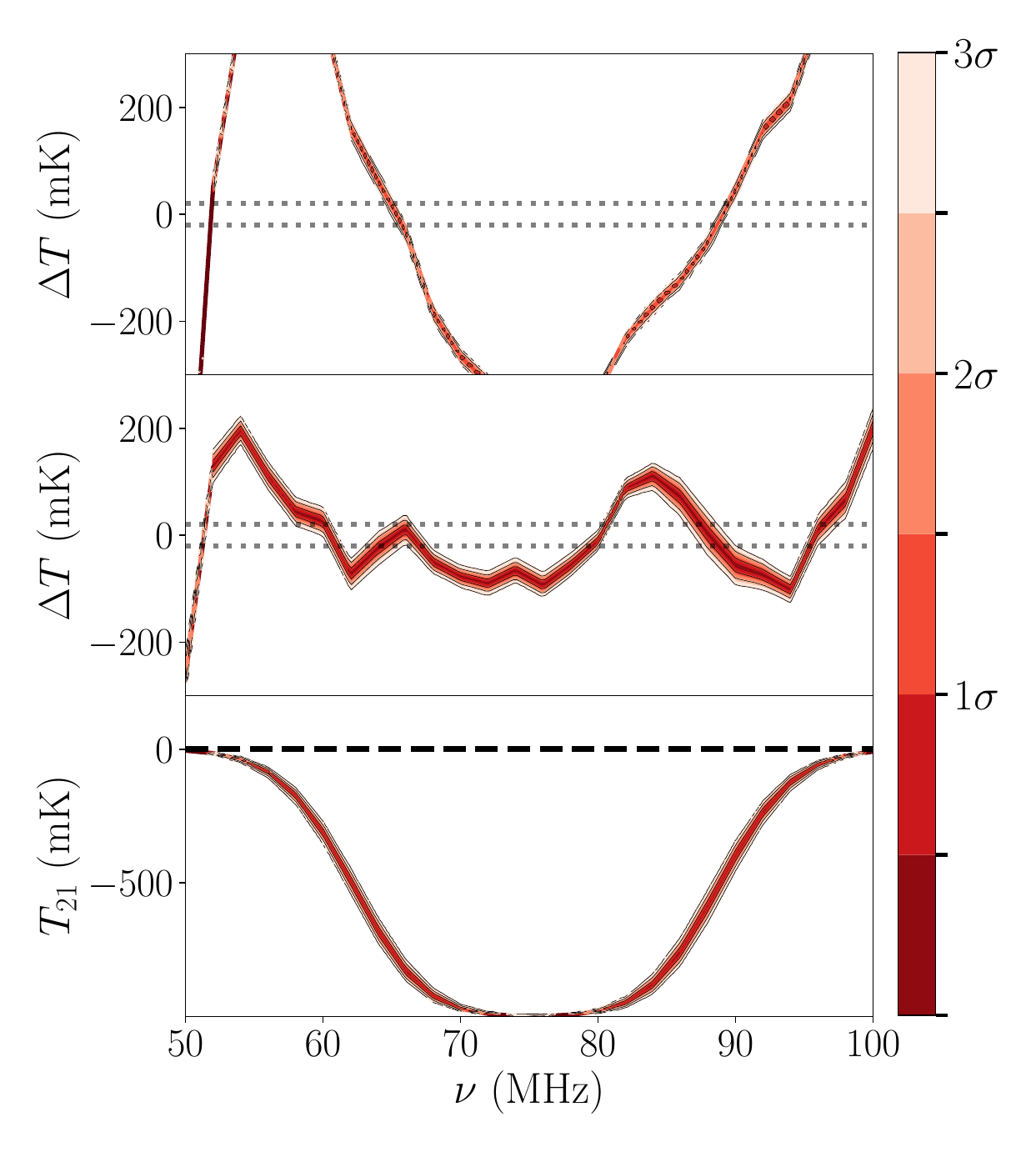}
        \end{subfigure}
        }
        \caption{
            Validation test results. Posterior predictive densities of the foreground fit residuals $\bm{r}_{i\mathrm{b}} =  [\bm{D}_\mathrm{v} - \bm{M}_{i\mathrm{b}}(\sTheta_{i\mathrm{b}})]$ (top panels), of the composite model's fit residuals $\bm{r}_{i\mathrm{c}} =  [\bm{D}_\mathrm{v} - \bm{M}_{i\mathrm{c}}(\sTheta_{i\mathrm{c}})]$ (middle panels) and of the model of signal (A) recovered with the composite model ($\bm{M}_\mathrm{a}(\sTheta_{i\mathrm{a}})$; bottom panels) for model classes $i=1$, 2 and 3 (subfigures a, b and c, respectively). The dotted lines in the top and middle panels denote the noise level in the simulated data. The dashed black line shows the input 21-cm signal, $\bm{S}_\mathrm{a}$, in the mock data. The subfigure captions indicate whether the model class passed model validation (PV) or failed (FV).
    }
\label{Fig:ValidationSignalRecoveryPosteriorPDs}
\end{figure*}

The null test results can be understood intuitively by examining \Cref{Fig:ValidationSignalRecoveryPosteriorPDs}, which shows the posterior PDs of the foreground fit residuals to the validation data $\bm{r}_{i\mathrm{b}} =  [\bm{D}_\mathrm{v} - \bm{M}_{i\mathrm{b}}(\sTheta_{i\mathrm{b}})]$ (top panels), of the composite model's fit residuals $\bm{r}_{i\mathrm{c}} =  [\bm{D}_\mathrm{v} - \bm{M}_{i\mathrm{c}}(\sTheta_{i\mathrm{c}})]$ (middle panels) and of $\bm{M}_\mathrm{a}(\sTheta_{i\mathrm{a}})$, the model of signal (A) recovered with the composite model (bottom panels) for composite/nuisance model sets $i=1$, 2 and 3 (\Cref{Fig:ValidationM1,Fig:ValidationM2,Fig:ValidationM3}, respectively). Here, $\sTheta_{i\mathrm{b}}$ and $\sTheta_{i\mathrm{c}}$ are the posterior distributions of the parameters of the $i$th foreground (signal (B)) model and composite 21-cm + foreground (signal (A) + signal (B)) model, respectively, derived from fits of those models to the data. $\sTheta_{i\mathrm{a}}$ is the marginal posterior distribution of the parameters of the $i$th 21-cm model derived in the fit of the $i$th composite model to the data.

Across the simulated spectral band, the posterior PDs of the residuals of both the $\bm{\mathcal{M}}_{1}$ foreground and the composite model fits to $\bm{D}_\mathrm{v}$ are consistent at 95\% credibility with the $20~\mathrm{mK}$ expected noise level in the data and each have a median RMS value of $17~\mathrm{mK}$. The posterior PD of the 21-cm signal associated with the fit of $\bm{M}_{1\mathrm{c}}$ has an amplitude consistent with zero across the band, implying its signal component is adding superfluous complexity in the context of this data set. The $\ln(\mathcal{B}_\mathrm{cb}^\mathrm{v}) = -3.2$ log-Bayes-factor found disfavouring $\bm{M}_{1\mathrm{c}}$ relative to $\bm{M}_{1\mathrm{b}}$ is qualitatively consistent with these results.

The posterior PDs of the foreground fit residuals of $\bm{M}_{2\mathrm{b}}$ to $\bm{D}_\mathrm{v}$ have a median RMS value of $\mathrm{RMS}_\mathrm{med}(\bm{r}_{2\mathrm{b}}) = 74~\mathrm{mK}$ and thus are inconsistent with the expected noise level. In contrast, the posterior PDs of the fit residuals of $\bm{M}_{2\mathrm{c}}$ to $\bm{D}_\mathrm{v}$ are consistent at 95\% credibility with the expected noise level in the data across the band and have a median RMS value of $\mathrm{RMS}_\mathrm{med}(\bm{r}_{2\mathrm{c}}) = 18~\mathrm{mK}$. The posterior PD of the 21-cm signal associated with the fit of $\bm{M}_{2\mathrm{c}}$ to the validation data has a mean amplitude $A = 381 \pm 36~\mathrm{mK}$ and central frequency at $\nu_{0} = 74 \pm 1~\mathrm{MHz}$, corresponding to a high significance spurious detection of the signal in the validation data. Relative to $\bm{M}_{2\mathrm{b}}$, the significantly improved fit of $\bm{M}_{2\mathrm{c}}$ to $\bm{D}_\mathrm{v}$ and the fact that the amplitude of the signal component of the model is highly inconsistent with zero ($\sim 10 \sigma$) are qualitatively consistent with the $\ln(\mathcal{B}_\mathrm{cb}^\mathrm{v}) = 184.6$ log-Bayes-factor found in favour of $\bm{M}_{2\mathrm{c}}$ over $\bm{M}_{2\mathrm{b}}$.

The posterior PDs of both the $\bm{\mathcal{M}}_{3}$ foreground and composite model fit residuals to the validation data have median RMS values of $294$ and $85~\mathrm{mK}$, respectively; thus both are inconsistent with the expected noise level in the data. The posterior PD of the 21-cm signal associated with the fit of $\bm{r}_{3\mathrm{c}}$ to $\bm{D}_\mathrm{v}$ has a mean amplitude $A = 998 \pm 2~\mathrm{mK}$ and central frequency at $\nu_{0} = 75.5 \pm 0.1~\mathrm{MHz}$, corresponding to a very high significance spurious detection of the signal in the validation data. Despite its poor fit in absolute terms, the improvement in the fit of $\bm{M}_{2\mathrm{c}}$ to $\bm{D}_\mathrm{v}$, relative to $\bm{M}_{2\mathrm{b}}$, and the fact that the amplitude of the signal component of the model is highly inconsistent with zero ($\sim 500 \sigma$) are qualitatively consistent with the $\ln(\mathcal{B}_\mathrm{cb}^\mathrm{v}) = 2997.2$ log-Bayes-factor in favour of $\bm{M}_{3\mathrm{c}}$ over $\bm{M}_{3\mathrm{b}}$.

\subsubsection{Euler set predictions}
\label{Sec:ModelsInABCABbarCABbarCbar}

To predict the Euler sets of the validated composite models we evaluate the predictivity and accuracy conditions given in \Cref{Eq:PredictivityCondition,Eq:AccuracyCondition} for the set of foreground models and the accuracy condition\footnote{We do not use the relative predictivities of the composite models for the foreground-only validation data as a useful metric for Euler set prediction since they suffer from the same limitation that necessitates the introduction of the BaNTER null test. Namely, BFBMC imposes an Occam penalty against a composite model composed of elements of $A$ and $B$ if the component $A$ represents unnecessary model complexity (as would be the case when fitting $\bm{D}_\mathrm{v}$). Thus, there is no guarantee that the aforementioned model will be more predictive than a less complex composite model composed of elements of $A$ and $\overline{B}$ that fits $\bm{D}_\mathrm{v}$ using spurious fits of its components.} for the composite models for $\bm{D}_\mathrm{v}$.

Assuming perfect validation data, foreground models that pass both conditions are elements of Euler set $B$. Assuming our signal model is an element of $A$, composite models that pass the accuracy condition are likely but not guaranteed to be elements of Euler set $C$. The lack of guarantee derives from the fact that while composite models with components drawn from $A$ and $B$ will necessarily be elements of $ABC$ those with components drawn from $A$ and $\overline{B}$ may be elements of $A\overline{B}C$ with respect to $\bm{D}_\mathrm{v}$ while being elements of $A\overline{B}~\overline{C}$ with respect to true data. Specifically, this difference in model classification with respect to the validation and true data will occur if structure in the validation data that is unmodellable by the foreground model can be absorbed by the signal model but the sum of this structure and the true signal in the data cannot.

The results of the predictivity and accuracy conditions for the foreground and composite models are summarised in \Cref{Tab:ForegroundPredictivityAccuracy}. We find that $\bm{M}_{1\mathrm{b}}$ is an element of $B$ and $\bm{M}_{2\mathrm{b}}$ and $\bm{M}_{3\mathrm{b}}$ are elements $\overline{B}$. Calculating the corresponding accuracy metrics for $\bm{M}_{1\mathrm{c}}$, $\bm{M}_{2\mathrm{c}}$ and $\bm{M}_{3\mathrm{c}}$, we find that $\bm{M}_{1\mathrm{c}}$ and $\bm{M}_{2\mathrm{c}}$ pass the accuracy condition with $Q_{0.999}(\lambda_{1\mathrm{c}}) = 10$ and $Q_{0.999}(\lambda_{2\mathrm{c}}) = 4$ while $\bm{M}_{3\mathrm{c}}$ fails with $Q_{0.999}(\lambda_{3\mathrm{c}}) = -309$. Combining these results yields predictions of the Euler sets of the composite models for data of $ABC$, $A\overline{B}C$ and $A\overline{B}~\overline{C}$ for $\bm{M}_{1\mathrm{c}}$, $\bm{M}_{2\mathrm{c}}$ and $\bm{M}_{3\mathrm{c}}$, respectively.

\begin{table}
    \caption{
        Validation predictivity and accuracy condition results for the foreground models. $\mathcal{B}_{j\mathrm{max}} = \mathcal{Z}_{j}/\mathcal{Z}_{\mathrm{b,max}}$, with $\mathcal{Z}_{j}$ the Bayesian evidence of the $j$th foreground model for the validation data and $\mathcal{Z}_{\mathrm{b,max}}$ the highest Bayesian evidence foreground model for the validation data. $Q_{0.999}(\lambda_{j\mathrm{b}})$ is the accuracy condition (see \Cref{Sec:ModelAccuracyCondition}) evaluated for the fit of the $j$th foreground model to the validation data.
        }
    \centerline{
        \begin{tabular}{l l l}
            \hline
            Foreground model & $\log(\mathcal{B}_{j\mathrm{max}})$ & $Q_{0.999}(\lambda_{j\mathrm{b}})$ \\
            \hline
            $\bm{M}_{1\mathrm{b}}$ & 0 & 10 \\
            $\bm{M}_{2\mathrm{b}}$ & -194 & -188 \\
            $\bm{M}_{3\mathrm{b}}$ & -3331 & -3320 \\
            \hline
    \end{tabular}
}
\label{Tab:ForegroundPredictivityAccuracy}
\end{table}

\subsubsection{BaNTER validation results as model priors}

Given BaNTER null test results and the aforementioned classification predictions for $\bm{M}_{1\mathrm{c}}$, $\bm{M}_{2\mathrm{c}}$ and $\bm{M}_{3\mathrm{c}}$, we should anticipate the comparison of $\bm{M}_{1\mathrm{c}}$ and $\bm{M}_{3\mathrm{c}}$ being a \textit{category I} model comparison problem and comparison of $\bm{M}_{1\mathrm{c}}$ and $\bm{M}_{2\mathrm{c}}$ being a \textit{category II} model comparison problem.

In principle, given that both $\bm{\mathcal{M}}_{2}$ and $\bm{\mathcal{M}}_{3}$ have failed the validation null test, we would be justified in judging them not to be credible models with which one could recover unbiased estimates of the 21-cm signal in the mock data. This would leave $\bm{\mathcal{M}}_{1}$ as a decisively preferred set of models for the data. However, for the purpose of confirming the above predictions regarding the Euler set in which each model is an element and their model comparison categorisations, in the next section we perform a full Bayes-factor-based comparison of $\bm{M}_{1\mathrm{b}}$, $\bm{M}_{2\mathrm{b}}$, $\bm{M}_{3\mathrm{b}}$, $\bm{M}_{1\mathrm{c}}$, $\bm{M}_{2\mathrm{c}}$ and $\bm{M}_{3\mathrm{c}}$ as models for the mock data.

\subsection{Mock data results}
\label{Sec:BFBMCvsPOBMC}

\begin{table*}
    \caption{
        Mock data results. For each model class we list the Bayes factors in favour of a detection of $\bm{S}_\mathrm{a}$ in the mock data, $\ln(\mathcal{B}_\mathrm{cb})$, and between the highest Bayesian evidence composite model for the mock data and the remaining models, ($\ln(\mathcal{B}_{i\mathrm{max}})$). We also note the corresponding validation test results for each model class (see \Cref{Tab:lnBdetection1}) and the Euler set in which each composite model in each model set is an element, as inferred from combining the validation test results and $\ln(\mathcal{B}_{i\mathrm{max}})$. We note that the true Euler sets of the composite models confirm the predicted Euler sets in \Cref{Sec:SIMVAMPApplication} so we do not list them separately.
    }
    \centerline{
        \begin{tabular}{l l l l l }
            \hline
            Model class & $\ln(\mathcal{B}_\mathrm{cb})$ & $\ln(\mathcal{B}_{i\mathrm{max}})$ & Validation test  & True / predicted model \\
            & & & results & classifications \\
            \hline
            $\bm{\mathcal{M}}_{1}$ & 18.5 & -1.6 & Passed & $ABC$ \\
            $\bm{\mathcal{M}}_{2}$ & 523.5 & 0 & Failed & $A\overline{B}C$ \\
            $\bm{\mathcal{M}}_{3}$ & 3910.2 & -431.3 & Failed & $A\overline{B}~\overline{C}$ \\
            \hline
    \end{tabular}
}
\label{Tab:lnBdetection3}
\end{table*}

\subsubsection{Bayes factors, predictivity and accuracy}
\label{Sec:BayesFactorsAndSignalRecovery}

\Cref{Tab:lnBdetection3} lists values of $\ln(\mathcal{B}_\mathrm{cb})$ -- the Bayes factors between $\bm{M}_{i\mathrm{c}}$ and $\bm{M}_{i\mathrm{b}}$ as models for the mock data, $\bm{D}$ -- and $\ln(\mathcal{B}_{i\mathrm{max}})$ -- the Bayes factor between the highest Bayesian evidence composite model for $\bm{D}$ and the remaining models.

We find a decisive preference ($\ln(\mathcal{B}_\mathrm{cb}) \geq 5$) for $\bm{M}_{i\mathrm{c}}$ over $\bm{M}_{i\mathrm{b}}$ in all three composite/nuisance model sets. In the absence of validation results, this implies decisive detections of either $\bm{S}_\mathrm{a}$ or systematic structure fit by the 21-cm model\footnote{Models $\bm{M}_{2\mathrm{c}}$ and $\bm{M}_{3\mathrm{c}}$ failed validation, indicating these models will yield biased 21-cm signal inferences. As such, the detections of $\bm{S}_\mathrm{a}$ with these models cannot be treated as definitive. Only $\bm{M}_{1\mathrm{c}}$, which passes validation, provides robust evidence for $\bm{S}_\mathrm{a}$ in the data.}.

Comparing models with BFBMC alone yields $\bm{M}_{2\mathrm{c}}$ as the highest Bayesian evidence model for the data. However, the Bayes factor $\ln(\mathcal{B}_{12}) = -1.6$ implies only a moderate preference for $\bm{M}_{2\mathrm{c}}$ over $\bm{M}_{1\mathrm{c}}$, while $\bm{M}_{3\mathrm{c}}$ is decisively disfavoured relative to both.

Calculating the accuracy condition given by \Cref{Eq:AccuracyCondition} for $\bm{M}_{1\mathrm{c}}$, $\bm{M}_{2\mathrm{c}}$ and $\bm{M}_{3\mathrm{c}}$ as models for the mock data, we find $\bm{M}_{1\mathrm{c}}$ and $\bm{M}_{2\mathrm{c}}$ pass with $Q_{0.999}(\lambda_\mathrm{1c}) = 9$ and $Q_{0.999}(\lambda_\mathrm{2c}) = 7$ and $\bm{M}_{3\mathrm{c}}$ fails with $Q_{0.999}(\lambda_\mathrm{3c}) = -412$. Thus, $\bm{M}_{3\mathrm{c}}$ is an element of $\overline{C}$ and, consistent with its low evidence, is not a credible model with which unbiased inferences of the 21-cm signal can be made. In contrast, $\bm{M}_{1\mathrm{c}}$ and $\bm{M}_{2\mathrm{c}}$ pass the predictivity and accuracy conditions given by \Cref{Eq:PredictivityCondition,Eq:AccuracyCondition}. In the absence of the BaNTER validation analysis this would limit them to being elements either of $ABC$ or $A\overline{B}C$, and, thus, it does not guarantee that unbiased inferences of the 21-cm signal are recoverable using them.

\begin{figure*}
    \centerline{
        \begin{subfigure}[t]{0.33\textwidth}
            \caption{\Large{$\bm{\mathcal{M}}_{1}$, PV}}
            \label{Fig:M1}
            \includegraphics[width=\textwidth]{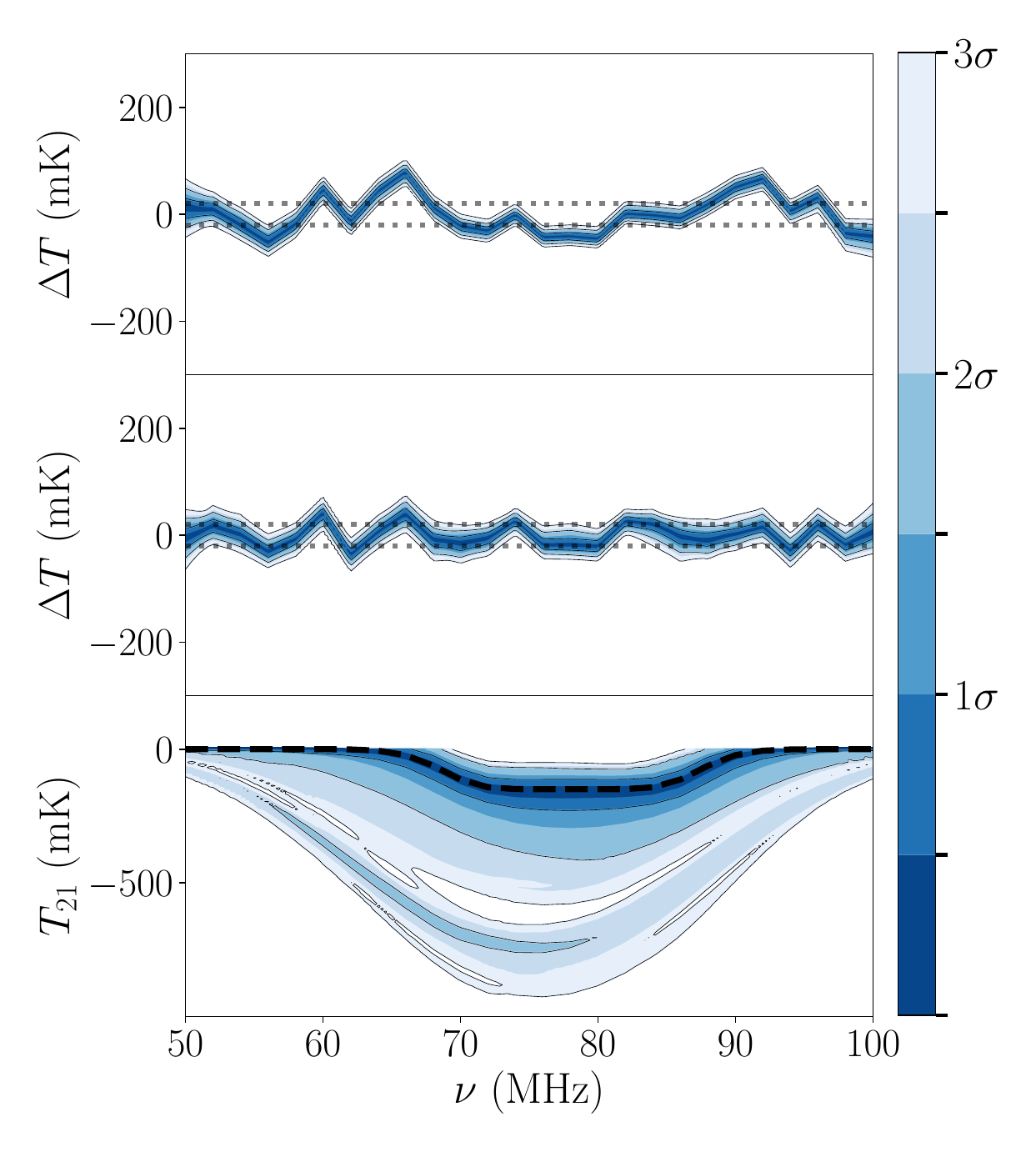}
        \end{subfigure}
        \begin{subfigure}[t]{0.33\textwidth}
            \caption{\Large{$\bm{\mathcal{M}}_{2}$, FV}}
            \label{Fig:M2}
            \includegraphics[width=\textwidth]{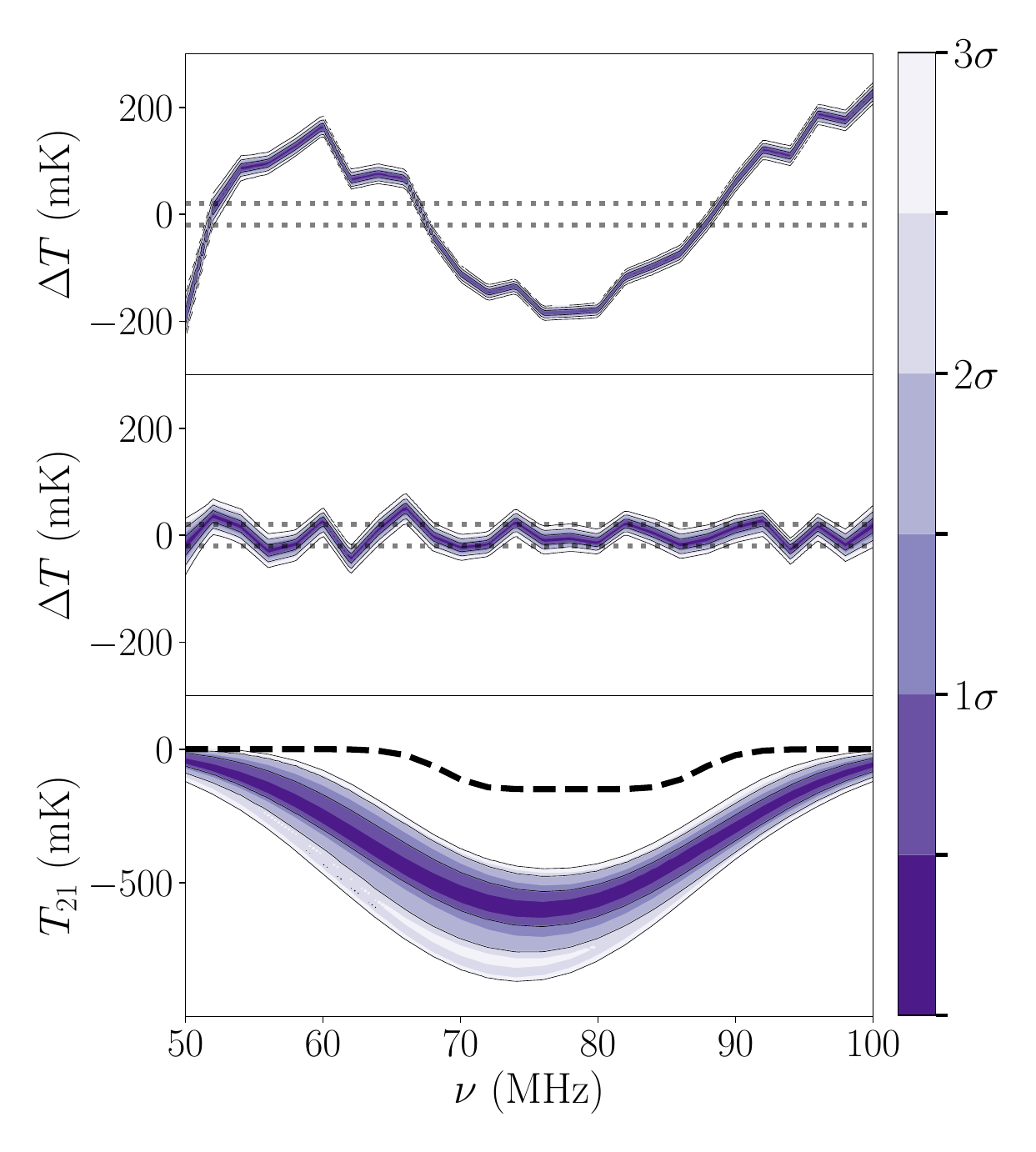}
        \end{subfigure}
        \begin{subfigure}[t]{0.33\textwidth}
            \caption{\Large{$\bm{\mathcal{M}}_{3}$, FV}}
            \label{Fig:M3}
            \includegraphics[width=\textwidth]{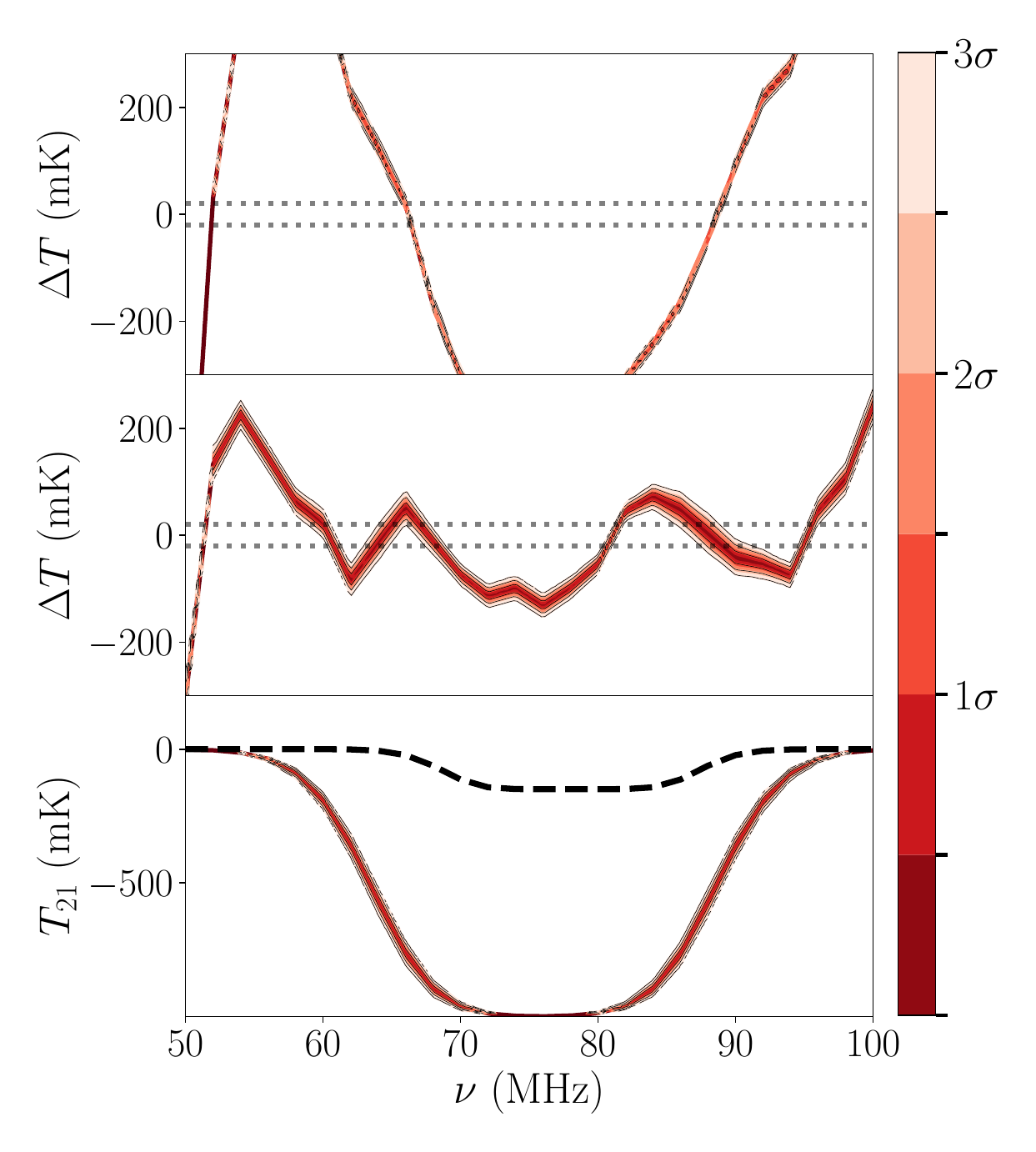}
        \end{subfigure}
        }
        \caption{
            Mock data results. Posterior predictive densities of the foreground fit residuals $\bm{r}_{i\mathrm{b}} =  [\bm{D} - \bm{M}_{i\mathrm{b}}(\sTheta_{i\mathrm{b}})]$ (top panels), composite models fit residuals $\bm{r}_{i\mathrm{c}} =  [\bm{D} - \bm{M}_{i\mathrm{c}}(\sTheta_{i\mathrm{c}})]$ (middle panels) and of the model of signal (A) recovered with the composite model ($\bm{M}_\mathrm{a}(\sTheta_{i\mathrm{a}})$; bottom panels) for model classes $i=1$, 2 and 3 (subfigures a, b and c, respectively). The dotted lines in the top and middle panels denote the noise level in the simulated data. The dashed black line shows the input 21-cm signal, $\bm{S}_\mathrm{a}$, in the mock data. The subfigure captions indicate whether the model class passed model validation (PV) or failed (FV).
    }
\label{Fig:SignalRecoveryPosteriorPDs}
\end{figure*}

\subsubsection{Signal recovery using Bayesian workflow with uninformative model priors}
\label{Sec:SignalRecoveryInUnvalidatedWorkflow}

\Cref{Fig:SignalRecoveryPosteriorPDs} shows posterior PDs derived from fits of the models to the mock data. Analysis of these results shows the following:
\begin{enumerate}
    \item $\bm{M}_{1\mathrm{c}}$: Accurately fits the mock data (\Cref{Fig:M1}, middle panel) and yields unbiased inferences of the 21-cm signal (\Cref{Fig:M1}, bottom panel).
    \item $\bm{M}_{2\mathrm{c}}$: Accurately fits the mock data (\Cref{Fig:M2}, middle panel) but results in biased inferences of the 21-cm signal (\Cref{Fig:M2}, bottom panel).
    \item $\bm{M}_{3\mathrm{c}}$: Fails to fit the mock data (\Cref{Fig:M3}, middle panel) and produces biased inferences of the 21-cm signal (\Cref{Fig:M3}, bottom panel).
\end{enumerate}

The posterior odds of $\bm{M}_{1\mathrm{c}}$, $\bm{M}_{2\mathrm{c}}$, and $\bm{M}_{3\mathrm{c}}$ are determined by the product of prior odds and Bayes factors between them. Under the \textit{Bayesian workflow with uninformative model priors} (\Cref{Fig:BayesianInferenceFlowChart}), equal prior odds are assumed, and BFBMC alone is used to calculate model weights based on observational data. This assigns approximately $83\%$ weight to $\bm{M}_{2\mathrm{c}}$ and $17\%$ to $\bm{M}_{1\mathrm{c}}$, with $\bm{M}_{3\mathrm{c}}$ contributing negligibly (see \Cref{Eq:BayesEqnForModels} and \Cref{Tab:aPosterioriPreference}). The high weighting of $\bm{M}_{2\mathrm{c}}$ results in significantly biased Bayesian model-averaged 21-cm signal inferences (with the averaged 21-cm signal being similar to \Cref{Fig:M2}, bottom panel).

The limitation of BFBMC-alone arises because it compares model predictivity for the data in aggregate. This does not distinguish between a model in $ABC$, with a high Bayesian evidence due to accurate fits of the model components, as with $\bm{M}_{1\mathrm{c}}$, and a model in $A\overline{B}C$, with biased fits of the model components that still produce good overall predictions, as with $\bm{M}_{2\mathrm{c}}$.

\subsubsection{Signal recovery using BaNTER-validated Bayesian inference workflow}
\label{Sec:SignalRecoveryInBaNTERvalidatedWorkflow}

Following the \textit{Model-validated Bayesian inference workflow} (\Cref{Fig:BayesianInferenceFlowChart}), we update the prior odds we assign to the models based on the results of the BaNTER validation. Both $\bm{M}_{2\mathrm{c}}$ and $\bm{M}_{3\mathrm{c}}$ fail the BaNTER null test, indicating that their inferred 21-cm signals are likely to be biased even if they achieve a high Bayesian evidence and/or accurate fit to the data in aggregate (as is the case with $\bm{M}_{2\mathrm{c}}$). As can be seen from \Cref{Fig:M2,Fig:M3}, the results of the mock data analysis confirm this to be the case.

In contrast, $\bm{M}_{1\mathrm{c}}$ passes the accuracy and predictivity conditions \textit{and} BaNTER validation, indicating that unbiased estimates of the 21-cm signal are recoverable with this model. The results of the mock data analysis in \Cref{Fig:M1} confirm the predicted result, recovering a posterior PD of the 21-cm signal consistent with the true 21-cm signal in the data.

Since $\bm{M}_{2\mathrm{c}}$ and $\bm{M}_{3\mathrm{c}}$ fail BaNTER validation, their prior odds can be approximated as negligible, leaving $\bm{M}_{1\mathrm{c}}$ as the only credible model for unbiased signal estimation. Thus, the ultimate result of the \textit{Model-validated Bayesian inference workflow} is the recovery of unbiased 21-cm signal inferences with $\bm{M}_{1\mathrm{c}}$ (\Cref{Fig:M1}, bottom panel).

\subsubsection{Comparison of mock data results and validation analysis predictions}
\label{Sec:CategoryIvsII}

From the results in \Cref{Sec:BayesFactorsAndSignalRecovery,Sec:SignalRecoveryInUnvalidatedWorkflow,Sec:SignalRecoveryInBaNTERvalidatedWorkflow} one can conclude the following:
\begin{enumerate}
    \item $\bm{M}_{2\mathrm{b}}$ and $\bm{M}_{3\mathrm{b}}$ are insufficiently accurate models of $\bm{S}_{\mathrm{b}}$ to yield unbiased estimates of the 21-cm signal when they are jointly fit to the data with $\bm{M}_{\mathrm{a}}$,
    \item in contrast, $\bm{M}_{1\mathrm{b}}$  is a sufficiently accurate model of $\bm{S}_{\mathrm{b}}$ to yield unbiased estimates of the 21-cm signal when jointly fit to the data with $\bm{M}_{\mathrm{a}}$,
    \item the true model classifications of $\bm{M}_{1\mathrm{c}}$, $\bm{M}_{2\mathrm{c}}$ and $\bm{M}_{3\mathrm{c}}$ are $ABC$, $A\overline{B}C$ and $A\overline{B}~\overline{C}$, respectively,
    \item comparison of $\bm{M}_{1\mathrm{c}}$ and $\bm{M}_{3\mathrm{c}}$ is a \textit{category I} model comparison problem that is successfully solved by the \textit{Bayesian workflow with uninformative model priors} (see \Cref{Fig:BayesianInferenceFlowChart}),
    \item comparison of $\bm{M}_{1\mathrm{c}}$ and $\bm{M}_{2\mathrm{c}}$ is a \textit{category II} model comparison problem. In this case, the \textit{Bayesian workflow with uninformative model priors} yields biased inferences of $\bm{S}_\mathrm{a}$, whereas unbiased inferences of $\bm{S}_\mathrm{a}$ are recovered by using the BaNTER validation results in the \textit{model-validated Bayesian inference workflow}.
\end{enumerate}
These observations are fully consistent with the expected outcomes based on the results of the validation analysis (see \Cref{Sec:SIMVAMPApplication}).

%%%%%%%%%%%%%%%%%%%%%%%%%%%%%%%%%%%%%%%%%%%%%%%%%%
\section{Discussion}
\label{Sec:Discussion}
%%%%%%%%%%%%%%%%%%%%%%%%%%%%%%%%%%%%%%%%%%%%%%%%%%

\subsection{Nested models}
\label{Sec:NestedVsNonNestedModels}

The flexible nature and utility of nested or partially nested models has led to their widespread usage in various fields spanning psychology and finance (e.g. \citealt{neco.2008.12-06-420, FAMA19933}) to ecology and astronomy (e.g. \citealt{annurev.ecolsys.110308.120159, 2016MNRAS.462.3069S, 2019MNRAS.488.2904S, 2022MNRAS.517..910S, 2022MNRAS.517..935S, 2023MNRAS.520.4443B}). Their ability to account for the a priori unknown complexity of foreground emission and uncertain presence of systematics has also led to their use in global 21-cm cosmology (e.g. B18, H18 \citealt{2018ApJ...853..187T, 2020ApJ...905..113H, 2020ApJ...897..174R, 2020MNRAS.492...22S, 2021MNRAS.506.2041A, 2022MNRAS.515.4565S}; S23, \citealt{2023MNRAS.522.1022S, 2024MNRAS.527.5649P, 2024MNRAS.527.2413P}).

The set of models, $\bm{\mathcal{M}} = \{\bm{M}_{1\mathrm{c}}, \bm{M}_{1\mathrm{b}}, \bm{M}_{2\mathrm{c}}, \bm{M}_{2\mathrm{b}}, \bm{M}_{3\mathrm{c}}, \bm{M}_{3\mathrm{b}}\}$, considered in this work are nested. Specifically, the parameters of $\bm{M}_{1\mathrm{b}}, \bm{M}_{2\mathrm{c}}, \bm{M}_{2\mathrm{b}}, \bm{M}_{3\mathrm{c}}$ and $\bm{M}_{3\mathrm{b}}$ are a subset of the parameters of $\bm{M}_{1\mathrm{c}}$ and the parameters of $\bm{M}_{2\mathrm{b}}$ and $\bm{M}_{3\mathrm{b}}$ are a subset of the parameters of $\bm{M}_{1\mathrm{b}}$. Thus, given that we use $\bm{M}_{1\mathrm{c}}$ as the generative model for the data, the nested structure of the other two nuisance models mean that $\bm{M}_{2\mathrm{b}}$ and $\bm{M}_{3\mathrm{b}}$ will, in general, provide incomplete descriptions of the validation data. Thus, in some respects, one may find it unsurprising that $\bm{\mathcal{M}}_{2}$ and $\bm{\mathcal{M}}_{3}$ fail BaNTER validation. However, we note that $\bm{\mathcal{M}}_{2}$ and $\bm{\mathcal{M}}_{3}$ will fail BaNTER validation only if the errors due to the imperfect modelling of the validation data by $\bm{M}_{2\mathrm{b}}$ and $\bm{M}_{3\mathrm{b}}$ are accurately modellable by $\bm{M}_{\mathrm{a}}$. Thus, their failure is not a foregone conclusion; rather, it derives from the properties of the $\bm{M}_{1\mathrm{b}}$ and $\bm{M}_{1\mathrm{c}}$ model incompletenesses and the structure of $\bm{M}_{\mathrm{a}}$, in combination.

Additionally, we note that while nested models were selected to construct illustrative \textit{category I} and \textit{category II} model comparison problems in this work, \textit{category II} model comparison problems will arise whenever the set of models under consideration include elements of both $ABC$ and $A\overline{B}C$ (see \Cref{Sec:CategoryII}), regardless of whether the nuisance components of the models are nested. \textit{Category II} model comparison problems with non-nested nuisance structure models are considered further in Sims et al. (in preparation).

\subsection{Other modelling scenarios - overly complex models that pass model validation}
\label{Sec:OverlyComplexModels}

In this work we used $\bm{\mathcal{M}}_{1}$ -- the most complex of the three model classes under consideration -- as our generative model for the mock data, with $\bm{\mathcal{M}}_{2}$ and $\bm{\mathcal{M}}_{3}$ containing a subset of the structural components present in $\bm{\mathcal{M}}_{1}$. A question of interest, not answered by the analysis considered thus far is: how would models that have a superset of the components of $\bm{\mathcal{M}}_{1}$ behave in the validation and data analysis?

Intuitively, one would expect such models to pass model validation but to be disfavoured at the data analysis stage. This disfavouring occurs due to the Occam penalty associated with the superfluous complexity and reduced predictivity of said models not being compensated by an improved fit to the data. To test this scenario, we consider an $\bm{\mathcal{M}}_{4}$ model class containing models defined identically to those in $\bm{\mathcal{M}}_{1}$ (see \Cref{Sec:SimulatedDataAndDataModels}) except that the foreground component $\bm{M}_{4\mathrm{b}}$ has an additional instrument-foreground coupling term ($N_\mathrm{pert} = 4$) relative to the generative model. Results derived from adding this model to the validation and data model comparison analyses conducted in \Cref{Sec:SIMVAMPApplication,Sec:BFBMCvsPOBMC} confirm the above prediction. Specifically, we find the following.
\begin{itemize}
    \item $\bm{\mathcal{M}}_{4}$ passes the model validation test with $\ln(\mathcal{B}_\mathrm{cb}^\mathrm{v}) = -3.5$, which corresponds to a strong preference (see \Cref{Tab:aPosterioriPreference}) for $\bm{M}_{4\mathrm{b}}$, relative to $\bm{M}_{4\mathrm{c}}$, as a model for the validation data.
    \item A fit of $\bm{M}_{4\mathrm{c}}$ to $\bm{D}$ yields a comparable recovery of the 21-cm signal to that obtained with $\bm{M}_{1\mathrm{c}}$. However, BFBMC of the models for the data yields $\ln(\mathcal{B}_{4\mathrm{max}}) = -3.7$ and $\ln(\mathcal{B}_{41}) = -2.1$. Thus, $\bm{M}_{4\mathrm{c}}$ is strongly and moderately disfavoured relative to $\bm{M}_{2\mathrm{c}}$ and $\bm{M}_{1\mathrm{c}}$, respectively, with the latter statistic representing the Occam penalty associated with the superfluous complexity of the nuisance structure component of the model.
\end{itemize}

\subsection{Model validation challenges and mitigation approaches}
\label{Sec:ModelValidationChallenges}

\subsubsection{Validation data fidelity}
\label{Sec:ValidationDataFidelity}

The efficacy of the model validation methodology introduced in this work is dependent on the fidelity of the model validation data. As such, if the nuisance component of the data is, for example, time-dependent\footnote{In 21-cm cosmology time-dependence of the astrophysical foregrounds, ionospheric conditions, ambient conditions at the observation site and the instrumental calibration may all contribute to the time-dependence of the data.}, care must be taken to ensure acquisition of the $\bm{S}_\mathrm{b}$-only observations required for $\bm{D}_\mathrm{v}$ and the observational data, $\bm{D}$, in a concomitant manner. Similarly, if it is not experimentally feasible to obtain $\bm{S}_\mathrm{b}$-only observations for $\bm{D}_\mathrm{v}$ and simulated $\bm{S}_\mathrm{b}$-only observations are used instead, the efficacy of the validation analysis will be a function of the accuracy of the simulated observations.

\subsubsection{Simulated validation data}
\label{Sec:SimulatedValidationData}

When validation simulations imperfectly represent the nuisance structure in the data, two scenarios must be considered: either there may be additional structure in the data not present in the simulations, or the simulations may contain additional structure not present in the data.

If the simulation has additional structure not present in the data models that are sufficient to describe the true nuisance structure in the data may fail the validation test. This could result in less stringent constraints on the SOI due to the nuisance structure model that pass validation being unnecessarily complex. However, these constraints will still be unbiased since models that pass the validation analysis with additional nuisance structure can also be expected to pass the simpler validation null test that would be associated with a more accurate validation data set.

Conversely, if the data contains additional nuisance structure not modelled in the validation simulations, a model that passes validation with the simulated data but fails to describe the additional structure might cause biased final inferences despite passing the validation test. The Bayesian averaged posterior inference resulting from the \textit{model-validated Bayesian inference workflow} will, nevertheless, be improved relative to those from the \textit{Bayesian workflow with uninformative model priors} by the downweighting or removal of the models that fail the validation null test.

\subsubsection{Accounting for uncertainties in validation data simulations}
\label{Sec:MitigatingRisksAssociatedWithImperfectValidationData}

Most often, simulations are a simplified version of nature; thus, the second case considered in \Cref{Sec:SimulatedValidationData} is more likely than the first. If the primary source of errors in the simulation derives from uncertainties in the correct specification of its input parameters (for example, due to the finite precision with which those parameters can be experimentally constrained), the validation methodology proposed in this work can be updated as follows. To account for the range of structure potentially present in the nuisance component of the data, one can replace the single validation data set considered thus far with a testing set of validation simulations spanning the range of possible nuisance structure consistent with the uncertainties on its input parameters. In this regime, the unperturbed simulation represents a fiducial validation data set that describes the most probable representation of the nuisance structure in the data. However, now, rather than using the fiducial validation data alone, for each model the validation analysis is iterated over all validation data sets. In so doing, the updated validation analysis accounts for other possible realisations of the nuisance structure in the data.

In this updated framework, one might conservatively discard any model that fails validation for any validation data set. Alternatively, if the values of input parameters for the validation data are drawn from a prior distribution one may calculate a prior probability associated with a given validation data set. This prior can then be used to either,
\begin{enumerate}
    \item inform the probability one places in a model that passes or fails the validation analysis with that simulated data, or,
    \item  update the validation threshold, $\ln(\mathcal{B}_\mathrm{threshold}^\mathrm{v})$, above which one considers the model to have failed the validation test\footnote{For example, if the perturbations in the simulation of the validation data are relatively improbable one may use a larger threshold $\ln(\mathcal{B}_\mathrm{threshold}^\mathrm{v})$ for the validation null-test or, for a fixed threshold, assign a model that narrowly fails this iteration of the updated validation analysis a reduced but non-zero probability assuming it passes other iterations.}.
\end{enumerate}

%%%%%%%%%%%%%%%%%%%%%%%%%%%%%%%%%%%%%%%%%%%%%%%%%%
\section{Summary \& Future work}
\label{Sec:Conclusions}
%%%%%%%%%%%%%%%%%%%%%%%%%%%%%%%%%%%%%%%%%%%%%%%%%%

In this paper we introduced the concept of \textit{category I} and \textit{category II} model comparison problems. A \textit{category I} scenario arises when all predictive composite models for the data also have predictive component models. In contrast, a \textit{category II} scenario involves a subset of models that can accurately fit the data but only with biased component models.

Bayes-factor-based model comparison (BFBMC) is a common Bayesian approach to comparing competing models for a data set. The comparison of competing models using BFBMC enables one to identify preferred models for the data in a \textit{category I} model comparison scenario. However, we demonstrated that in a \textit{category II} model comparison scenario, it will favour predictive models for the data, regardless of whether the components of those models are accurate.

To address the challenges of deriving unbiased inferences in \textit{category II} scenarios, we presented the Bayesian Null Test Evidence Ratio (BaNTER) validation framework. BaNTER calculates the Bayes factor between composite and single-component models for single-signal validation data. This allows one to identify, a priori, models in which a statistically significant improvement in the accuracy of their fit to the data is liable to result from biased fits of their model components. Single component models that are favoured over their composite variants are judged to have passed. Those for which the composite model is preferred, implying evidence in favour of a spurious detection of the absent signal component, fail.

We demonstrated the efficacy of this methodology for deriving unbiased inferences from a mock data set in a \textit{category II} model comparison scenario in a global 21-cm cosmology context. Comparing three competing composite models, $\bm{M}_{1\mathrm{c}}$, $\bm{M}_{2\mathrm{c}}$ and $\bm{M}_{3\mathrm{c}}$, for a data set where $\bm{M}_{1\mathrm{c}}$ is the generative model for the data, $\bm{M}_{2\mathrm{c}}$ accurately fits the data in aggregate but only with biased estimates of the foregrounds and 21-cm signal and $\bm{M}_{3\mathrm{c}}$ is a less predictive model of the data than $\bm{M}_{1\mathrm{c}}$ and $\bm{M}_{2\mathrm{c}}$, we demonstrated the following.
\begin{itemize}
    \item BFBMC successfully resolves the \textit{category I} problem by decisively favouring $\bm{M}_{1\mathrm{c}}$ over $\bm{M}_{3\mathrm{c}}$.
    \item The Bayesian evidence of $\bm{M}_{1\mathrm{c}}$ and $\bm{M}_{2\mathrm{c}}$ are comparable because each provides a comparable fit to the data over their respective prior volumes. Consequently, a \textit{Bayesian workflow with uninformative model priors} in which BFBMC alone is used to derive the posterior odds of the models, fails to solve the \textit{category II} model comparison problem of identifying $\bm{M}_{1\mathrm{c}}$ as superior to $\bm{M}_{2\mathrm{c}}$. Consequently, the Bayesian model averaged 21-cm signal inferences derived from this analysis are biased.
    \item Applying BaNTER validation to the three models classes we find that $\bm{M}_{1\mathrm{c}}$ passes. In contrast, $\bm{M}_{2\mathrm{c}}$ and $\bm{M}_{3\mathrm{c}}$ fail the null test due to spurious detections of a 21-cm signal in the foreground-only validation data.
    \item Incorporating the BaNTER validation results to derive informative model priors used in a \textit{model-validated Bayesian inference workflow} enables the derivation of unbiased inferences of the 21-cm signal in the data.
\end{itemize}

In Sims et al. (in preparation), we use BaNTER and the \textit{model-validated Bayesian inference framework}, developed in this work, as part of a broader comparison of a set of data models that have been applied to 21-cm signal estimation in the literature (\citealt{2018Natur.555...67B, 2018Natur.564E..32H, 2023MNRAS.521.3273S}). There it will be shown that selecting the best models for unbiased 21-cm signal inference from time-averaged spectrometer data constitutes a \textit{category II} model comparison problem, underscoring the necessity of model validation in this context. Application of BaNTER validation to other scientific problems involving the comparison of competing composite models for a data set provides a rich avenue for future work.

%%%%%%%%%%%%%%%%%%%%%%%%%%%%%%%%%%%%%%%%%%%%%%%%%%
\section*{Acknowledgements}
%%%%%%%%%%%%%%%%%%%%%%%%%%%%%%%%%%%%%%%%%%%%%%%%%%

This work was supported by the NSF through research awards for EDGES (AST-1813850, AST-1908933, and AST-2206766). PHS thanks Irina Stefan for valuable discussions and helpful comments on a draft of this manuscript. EDGES is located at the Inyarrimanha Ilgari Bundara, the CSIRO Murchison Radio-astronomy Observatory. We acknowledge the Wajarri Yamatji people as the traditional owners of the Observatory site. We thank CSIRO for providing site infrastructure and support.

%%%%%%%%%%%%%%%%%%%%%%%%%%%%%%%%%%%%%%%%%%%%%%%%%%
\section*{Data Availability}
%%%%%%%%%%%%%%%%%%%%%%%%%%%%%%%%%%%%%%%%%%%%%%%%%%

The data from this study will be shared on reasonable request to the corresponding author.



\bibliographystyle{mnras}
% \bibliography{example} % if your bibtex file is called example.bib
\bibliography{bibliography} % if your bibtex file is called bibliography.bib

%%%%%%%%%%%%%%%%%%%%%%%%%%%%%%%%%%%%%%%%%%%%%%%%%%

%%%%%%%%%%%%%%%%% APPENDICES %%%%%%%%%%%%%%%%%%%%%

\appendix

% %%%%%%%%%%%%%%%%%%%%%%%%%%%%%%%%%%%%%%%%%%%%%%%%%%
% \section{Some extra material}
% %%%%%%%%%%%%%%%%%%%%%%%%%%%%%%%%%%%%%%%%%%%%%%%%%%

%%%%%%%%%%%%%%%%%%%%%%%%%%%%%%%%%%%%%%%%%%%%%%%%%%
\section{Euler set notation in the context of 21-cm cosmology}
\label{Sec:EulerSets21-cmCosmologyExample}
%%%%%%%%%%%%%%%%%%%%%%%%%%%%%%%%%%%%%%%%%%%%%%%%%%

To provide intuition for the set notation introduced in \Cref{Sec:ModelPredictivityAndAccuracy}, in this appendix we consider how it applies to an example drawn from 21-cm cosmology, in which one aims to model data containing 21-cm radiation emitted by neutral hydrogen at Cosmic Dawn (the SOI; $\bm{S}_\mathrm{a}$) as well as nuisance radio frequency foreground emission and potential instrumental systematics ($\bm{S}_\mathrm{b}$; hereafter, foregrounds for brevity).

In the global 21-cm cosmology context, the 7 possible permutations of composite models described in \Cref{Sec:ComponentModelPermutations} are characterised by the  attributes summarised in \Cref{Tab:EulerSetsExample}.

\begin{table*}
    \caption{
        List of Euler sets permutations that composite models may be elements of, in the context of modelling a 21-cm cosmology data set containing two components: 21-cm radiation emitted by neutral hydrogen at Cosmic Dawn (the SOI; $\bm{S}_\mathrm{a}$) and nuisance foregrounds ($\bm{S}_\mathrm{b}$). We list for each permutation: whether the 21-cm and foreground component models associated with that Euler set are predictive and accurate of the 21-cm signal and foreground components of the data in isolation, whether the composite model is a predictive and accurate description of the data in aggregate, whether unbiased parameter inferences are recoverable when fitting a composite model in this Euler set to the data and whether model validation is necessary to distinguish between the composite model in the given Euler set and a model in the Euler set of interest, $ABC$.
    }
    \centerline{
        \begin{tabular}{l l l l l l }
            \hline
            Euler set permutation & Predictive and accurate & Predictive and accurate & Predictive and accurate & Unbiased parameter & Model validation  \\
            & 21-cm signal model & foreground model & composite model & inferences &  necessary \\
            \hline
            $ABC$ & \cmark & \cmark & \cmark & \cmark & - \\
            $\overline{A}BC$ & \xmark & \cmark & \cmark & \xmark & \cmark \\
            $A\overline{B}C$ & \cmark & \xmark & \cmark & \xmark & \cmark \\
            $\overline{A}~\overline{B}C$ & \xmark & \xmark & \cmark & \xmark & \cmark \\
            $A\overline{B}~\overline{C}$ & \cmark & \xmark & \xmark & \xmark & \xmark \\
            $\overline{A}B\overline{C}$ & \xmark & \cmark & \xmark & \xmark & \xmark \\
            $\overline{A}~\overline{B}~\overline{C}$ & \xmark & \xmark & \xmark & \xmark & \xmark \\
                \hline
    \end{tabular}
}
\label{Tab:EulerSetsExample}
\end{table*}

In detail, composite models in the Euler set $ABC$ are ideal models for the data with which unbiased estimates of the SOI can be obtained. These models are constructed by combining models for the 21-cm signal and foregrounds that are both accurate and predictive and the goal of composite model comparison is to identify models in this Euler set.
Composite models in $A\overline{B}C$, $\overline{A}BC$ and $\overline{A}~\overline{B}C$, in contrast, are more problematic. These models are capable of accurately fitting the data in aggregate, but \textit{only} via inaccuracies in the 21-cm model being absorbed by the foreground model or inaccuracies in the foreground model being absorbed by the 21-cm model or both of these effects occurring in unison, in the three sets respectively. In any of these cases, one will recover biased estimates of the SOI despite the composite model being predictive of the data in aggregate. Thus, when models of this sort are in the set of models for the data under consideration, model comparison using BFBMC alone will result in a degeneracy, in the space of a posteriori model probability, between these models and those in the set $ABC$. To eliminate this degeneracy and, with it, the bias in one's inferences of the SOI, identifying and removing models in the sets $A\overline{B}C$, $\overline{A}BC$ and $\overline{A}~\overline{B}C$ is essential. The BaNTER model validation framework introduced in \Cref{Sec:BaNTER} is designed for this purpose.

Models in $\overline{A}B\overline{C}$, $A\overline{B}~\overline{C}$ and $\overline{A}~\overline{B}~\overline{C}$ are inaccurate and/or not predictive in aggregate (see \Cref{Eq:PredictivityCondition,Eq:AccuracyCondition}) and are constructed by combining models for the 21-cm signal and foregrounds where the former, latter or both are inaccurate and/or not predictive. Unlike models in $A\overline{B}C$, $\overline{A}BC$ and $\overline{A}~\overline{B}C$, these models are identifiable and separable from those of interest, in $ABC$, using BFBMC of the composite models for the data alone. Thus, if the set of models under consideration derives exclusively from $ABC$, $\overline{A}B\overline{C}$, $A\overline{B}~\overline{C}$ and $\overline{A}~\overline{B}~\overline{C}$, BaNTER validation is incidental to recovery of unbiased inferences of the SOI.

In practice, the Euler set classifications of the set of models under consideration may be uncertain. In this case, BaNTER validation can be used to identify models that are likely to be biased when fit to the data, if present, and in so doing, facilitates unbiased inferences of the SOI following model comparison, regardless of the Euler set classifications of the models.

%%%%%%%%%%%%%%%%%%%%%%%%%%%%%%%%%%%%%%%%%%%%%%%%%%
\section{Category I and II model comparison in the context of 21-cm cosmology}
\label{Sec:CategoryIandII21-cmCosmologyExample}
%%%%%%%%%%%%%%%%%%%%%%%%%%%%%%%%%%%%%%%%%%%%%%%%%%

To provide intuition for the model comparison categories introduced in \Cref{Sec:CategoryI,Sec:CategoryII}, in this appendix we describe how they apply to the 21-cm cosmology example described in \Cref{Sec:EulerSets21-cmCosmologyExample}.

Suppose we have a global 21-cm signal data set, of the form described in \Cref{Eq:Data}, composed of two signal components -- 21-cm radiation emitted by neutral hydrogen at Cosmic Dawn (the SOI; $\bm{S}_\mathrm{a}$) and nuisance foregrounds ($\bm{S}_\mathrm{b}$) -- and noise.

Furthermore, we have a set of composite models for describing the data, $\bm{\mathcal{M}}_{\mathrm{c}}$. Each composite model is obtained by summing a model for the global 21-cm signal, $\bm{M}_{j\mathrm{a}}$ and a model for the foregrounds, $\bm{M}_{k\mathrm{b}}$, where these models are elements of the sets of models under consideration for these components: $\bm{\mathcal{M}}_{\mathrm{a}}$ and $\bm{\mathcal{M}}_{\mathrm{b}}$, respectively.

\textit{Category I} and \textit{II} global 21-cm composite model comparison problems, in the context of global 21-cm cosmology, are distinguished by whether the set of composite models for the data is composed exclusively of models for which unvalidated BFBMC is sufficient for identifying models that are both accurate and predictive models for the data \textit{and} accurate predictive component models for the 21-cm signal and foreground (those in Euler set $ABC$; see \Cref{Sec:EulerSets21-cmCosmologyExample}).

A \textit{category I} global 21-cm composite model comparison problem exists when unvalidated BFBMC is sufficient. This occurs when the elements of $\bm{\mathcal{M}}_{\mathrm{c}}$ are elements of the Euler sets $ABC$, $A\overline{B}~\overline{C}$, $\overline{A}B\overline{C}$ or $\overline{A}~\overline{B}~\overline{C}$. Here, composite models in the Euler set $ABC$ are composed of a 21-cm signal and foregrounds that are both accurate and predictive and fits of these models to the data enable unbiased estimates of the SOI to be obtained. In contrast, models in $\overline{A}B\overline{C}$, $A\overline{B}~\overline{C}$ and $\overline{A}~\overline{B}~\overline{C}$ are inaccurate and/or not predictive in aggregate and are constructed by combining models for the 21-cm signal and foregrounds where the former, latter or both are inaccurate and/or not predictive.

A \textit{category II} global 21-cm composite model comparison problem exists when unvalidated BFBMC is insufficient for uniquely identifying models in the Euler set $ABC$. This occurs when the elements of $\bm{\mathcal{M}}_{\mathrm{c}}$ are elements of the Euler sets $ABC$, and at least one of $\overline{A}BC$, $A\overline{B}C$ or $\overline{A}~\overline{B}~\overline{C}$. Composite models from the latter three Euler sets are capable of accurately fitting the data in aggregate, but \textit{only} via inaccuracies in the 21-cm model being absorbed by the foreground model or inaccuracies in the foreground model being absorbed by the 21-cm model or both of these effect occurring in unison. These models will yield biased estimates of the 21-cm signal when fit to the data but will be degenerate with models in $ABC$ in the space of model probability as judged by unvalidated BFBMC. In this case, identifying and removing models in these three Euler sets is essential and can be achieved using the BaNTER model validation framework introduced in \Cref{Sec:BaNTER}.

%%%%%%%%%%%%%%%%%%%%%%%%%%%%%%%%%%%%%%%%%%%%%%%%%%
\section{Foreground model description}
\label{Sec:ForegroundModelDescription}
%%%%%%%%%%%%%%%%%%%%%%%%%%%%%%%%%%%%%%%%%%%%%%%%%%

The individual terms in \Cref{Eq:BFCCdataModel} are defined as follows. $\bar{\bm{B}}_\mathrm{factor}(\bm{\nu})$ is an instrumental beam factor associated with the spectral structure of the spectrometer's directivity pattern. The first and second terms describe the spatially-isotropic and -anisotropic subcomponents of a power-law component of the foreground emission, respectively. The third term describes the beam-factor weighted cosmic microwave background (CMB) temperature and the spatially-isotropic subcomponent of the power-law component of the foreground emission, for which the beam-factor is not cancelled during a beam factor chromaticity correction (see S23). The common product of the terms in square brackets, $\e^{-\tau_\mathrm{ion}(\nu)}$, models ionospheric absorption. The fifth term models the beam-factor weighted net emission by hot electrons in the ionosphere.

\Cref{Eq:BFCCdataModel} has $N_\mathrm{pert} + 2$ astrophysical foreground parameters and 2 ionospheric foreground parameters. The free parameters of the astrophysical foreground model are defined as follows. $\bar{T}_\mathrm{m_{0}}$ describes the time- and sky-averaged foreground brightness temperature at reference frequency $\nu_\mathrm{c}$. $\beta_{0}$ describes the mean temperature spectral index of the power law component of the foreground emission. $p_{\alpha}$ describes the fractional amplitude of the $\alpha$th log-polynomial model vector for describing spectral fluctuations about the sky-averaged spectrum of the foreground brightness temperature field, normalised to the fractional amplitude of the perturbation relative to the mean brightness temperature at the reference frequency $\nu=\nu_\mathrm{c}$. The two free parameters of the ionospheric foreground model, $T_{\mathrm{e}}$ and $\tau_0$ describe the temperature of ionospheric electrons and the effective ionospheric optical depth at $\nu=\nu_\mathrm{c}$, respectively.

%%%%%%%%%%%%%%%%%%%%%%%%%%%%%%%%%%%%%%%%%%%%%%%%%%

% Don't change these lines
\bsp	% typesetting comment
\label{lastpage}
\end{document}